\pdfoutput=1

\documentclass[11pt]{article}
\usepackage[final]{acl}

\usepackage{times}
\usepackage{latexsym}
\usepackage[T1]{fontenc}
\usepackage[utf8]{inputenc}

\usepackage{algorithm}
\usepackage{algpseudocode}
\usepackage{float}

\usepackage{amsfonts}
\usepackage{amsmath}
\usepackage{amssymb}        
\usepackage{bold-extra}     
\usepackage{bm}             

\usepackage{graphicx}       
\usepackage{multirow}       
\usepackage{booktabs}       
\usepackage{enumitem}       
\usepackage{subcaption}     
\usepackage[hang,flushmargin]{footmisc}  

\definecolor{KDpurple}{rgb}{0.6,0.18,0.64}

\title{ECGNet: A generative adversarial network (GAN) approach to the synthesis of 12-lead ECG signals from single lead inputs
}

\author{  Max Bagga\thanks{These authors contributed equally to this work.} \\
  \small Department of Computer Science \\
  Emory University \\
  Atlanta, GA, 30322, USA \\
  \texttt{\small max.bagga@emory.edu} 
  \\\And
  Hyunbae Jeon\footnotemark[1] \\
  \small Department of Computer Science \\
  Emory University \\
  Atlanta, GA, 30322, USA \\
  \texttt{\small harry.jeon@emory.edu} \\\And
  Alex Issokson\footnotemark[1] \\
  \small Department of Computer Science \\
  Emory University \\
  Atlanta, GA, 30322, USA \\
  \texttt{\small alex.issokson@emory.edu} \\}
  
\begin{document}
\maketitle

\begin{abstract}
Electrocardiography (ECG) signal generation has been heavily explored using generative adversarial networks (GAN) because the implementation of 12-lead ECGs is not always feasible. The GAN models have achieved remarkable results in reproducing ECG signals but are only designed for multiple lead inputs and the features the GAN model preserves have not been identified—limiting the generated signals use in cardiovascular disease (CVD)-predictive models. This paper presents ECGNet which is a procedure that generates a complete set of 12-lead ECG signals from any single lead input using a GAN framework with a bidirectional long short-term memory (LSTM) generator and a convolutional neural network (CNN) discriminator. Cross and auto-correlation analysis performed on the generated signals identifies features conserved during the signal generation—i.e., features that can characterize the unique-nature of each signal and thus likely indicators of CVD. Finally, by using ECG signals annotated with the CVD-indicative features detailed by the correlation analysis as inputs for a CVD-onset-predictive CNN model, we overcome challenges preventing the prediction of multiple-CVD targets. Our models are experimented on 15s 12-lead ECG
dataset recorded using MyoVista’s wavECG. Functional outcome data for each patient is recorded and used in the CVD-predictive model. Our best GAN model achieves state-of-the-art accuracy with Fréchet Distance (FD) scores of 4.73, 4.89, 5.18, 4.77, 4.71, and 5.55 on the V1-V6 pre-cordial leads respectively and shows strength in preserving the P-Q segments and R-peaks in the generated signals. To the best of our knowledge, ECGNet is the first to predict all of the remaining eleven leads from the input of any single lead. Because of ECGNet's ability to conserve aspects of the PQRST complex, our work is useful for feature extraction that combats the black box nature of the GAN signal generation and ultimately future CVD-predictive models.
\end{abstract}
\section{Introduction}
\label{sec:introduction}
With cardiovascular disease (CVD) being the leading cause of death, understanding CVD predictors is vital \cite{roth-etal-2020-CVD-mortality}. Traditionally, cardiologists can use twelve-lead electrocardiography (ECG) signals to predict major CVD events near onset through the diagnosis of risk factors \cite{dawber-etal-1952-ecg-traditional-use}.
These risk factors are determined by observing abnormal electrical signal patterns produced by each lead in the ECG \cite{chiou-etal-2021-ecg-predictions}. However, these patterns tend to become recognizable to cardiologists only near inevitable CVD onset \cite{ebrahimi-etal-2020-cardiologist-shortcomings}. Consequently, through the availability of large data sets of patients' ECG signals, there has been a growing interest to use deep learning (DL) algorithms to extract features of ECG signals and use them to predict CVD before irreversible onset \cite{li-boulanger-2020-ecg-dl}. Despite the advancement in constructing DL-based CVD-predictive models, clinical implementation is challenging because of the models' inability to classify multiple-disease targets \cite{stracina-etal-2021-clinical}. \\
\indent 
CVD manifests into disorders that can be categorized: structural heart disease, functional heart disease, and hemodynamic disorders. Several studies have proposed DL models to overcome the multiple-disease target barrier by predicting CVDs within a single category. The convolutional neural network (CNN) rECHOmmend combines the outcomes of multiple structural diseases into a single prediction that outperforms single-disease target models \cite{ulloa-cerna-2022-rECHOmmend}. While rECHOmmend takes the indicators of multiple different structural CVDs and successfully predicts future onset of generic structural CVD, it fails to classify the onset of a specific structural CVD. Moreover, deep learning models predicting functional heart disease do not significantly outperform classical risk calculators, and those that are on par are single-disease target models—i.e. the models make predictions on just one functional heart disease \cite{zhou-etal-2021-functional-CVD}. Therefore, clinical implementation is still limited by the inability to specifically classify a disease or take multiple disease features as inputs \cite{stracina-etal-2021-clinical}. \citet{zhou-etal-2021-functional-CVD} argued that deep learning models could add unnecessary complexity and make it hard to investigate crucial feature inputs due to their black box nature. Applying this logic from \citet{zhou-etal-2021-functional-CVD}, because the true interactions between the input variables of rECHOmmend are obscured by the CNN's black box, the process cannot be extended to solve the functional heart disease models' inabilities to take multiple disease inputs and vice versa. \\
\indent 
In this paper, we present ECGNet: a Generative Adversarial Network (GAN) model with a bidirectional grid long short-term memory (LSTM) generator and a CNN discriminator trained to be able to reconstruct a complete twelve-lead ECG lead signal set from an input of any one ECG lead signal. The latent features conserved by the GAN model will be extracted from the reconstructed signals using signal cross-correlation and passed as input to another CNN model to predict the onset of multiple structural, functional, and hemodynamic CVDs. We hypothesize that our method can successfully classify multiple-disease targets because by first reconstructing ECG signals and identifying the conserved electrical motifs, we can understand the distinguishing ECG characteristics that the model actually uses, and thus the black-box process previously preventing transfer learning can be illuminated. Our approach uses a bidirectional LSTM-CNN GAN because one LSTM dimension can encode the signal's temporal trends while the other can encode the other features \cite{hazra-byun-2020-SynSigGAN}. \citet{zhu-etal-2019-bigridlstm} used a bidirectional LSTM-CNN GAN model to generate entire 12-signal data sets from patient features—a similar outcome as ours but with features instead of signals as inputs.   \\ 
\indent
Our approach is evaluated on ECG and functional outcome data from 976 patients with diverse racial, gender, physical, socioeconomic, and medical histories (Section \ref{sec:experiments}). The models operate on the 15-second MyoVista wavECG recordings \cite{sengupta-etal-myovista, Sengupta-etal-2018-wavecg} and over 2000 functional outcomes—e.g., biometrics and disease diagnosis—collected for each patient. We expect our experiments to show that our GAN architecture can recreate a full set of ECG signals from any single ECG lead input that the CNN discriminator cannot consistently distinguish from the actual signals. The signal cross-correlation analysis is expected to identify electric motifs that are conserved in intra-individual ECG signals. With the predictive model taking the conserved electric motifs as inputs, the onsets of CVDs are expected to be predicted and classified.\\ 
\indent 
This work makes three main contributions as follows:
\begin{enumerate}[leftmargin=*]
\item We train a GAN model to reconstruct the remaining eleven ECG lead signals from an input of any single ECG leads
\item We extract connections between each reconstructed signal's characteristics
\item The extracted features are used as inputs to predict and classify multiple CVDs
\end{enumerate}

To the best of our knowledge, this is the first work that reconstructs all twelve ECG signals from the input of any one lead signal.
\section{Related Work}
\label{sec:related-work}
The ECGNet procedure focuses on three aspects: reconstructing electrocardiography (ECG) lead signals, mathematical feature extraction from ECG lead signals, and deep learning methods for predicting cardiovascular disease (CVD). 
\subsection{Reconstructing ECG Lead Signals}
\label{ssec:reconstruction}
The task of reconstructing ECG lead signals is not a new endeavor. \citet{toyoshima-etal-1958-original-reconstruction} attempted to reconstruct the QRS ECG complexes by summing the ventricular activation or electrical potential changes of the heart in order to diagnose myocardial infarction; however, as demonstrated by \citet{oosterom-2002-solid-angle}, the accurate diagnosis of CVD by these methods cannot proceed without prior knowledge of the QRS complex trends corresponding to the CVD and performing the inverse operation on the ECG signal to identify the electrical potential changes of the heart. The mechanism proposed by \citet{toyoshima-etal-1958-original-reconstruction} is thus rendered obsolete as the QRS trends are what the mechanism attempts to establish yet is also required to make the mechanism accurate. This circular challenge persists throughout modern techniques as well and subsequently is a problem we attempt to circumnavigate by first using the GAN model (\ref{ssec:ganmodel}) to identify the latent motifs of ECG signal leads (\ref{ssec:cross-correlationanalysis}) and then using the motifs as inputs to the predictive model (\ref{ssec:cvd-predictivecnnmodel}). \\
\indent
\citet{drew-etal-2002-interpolation} was the first to construct multiple lead signals from the input of other lead signals. Using mathematical interpolation on a reduced lead data set, the four missing precordial lead signals were reconstructed in order to diagnose arrhythmia and myocardial ischemia. The reconstructed lead signals preserved the features of the CVDs. Consequently, if the interpolation is clinically implemented, the eight leads can be placed optimally instead of in their traditional twelve-lead placement. Despite the great work of \citet{drew-etal-2002-interpolation}, the interpolation cannot be generalized to even a particular category of CVDs as the eliminated leads of the analysis are invaluable in the diagnosis of other CVDs; therefore, with the advent of DL, many studies have sought to reconstruct ECG lead signals without losing the features of the eliminated leads. \\
\indent 
A major breakthrough was made when \citet{sohn-etal-2020-three-lead-dl} used a LSTM Network to reconstruct a full twelve-lead ECG set from the signal inputs of a portable three-lead ECG hardware. The network outperformed a three-lead input interpolation approach proposed by \citet{hsu-wu-2014-three-lead-inter} which already had state-of-the-art results proving the viability of long-term, reduced-lead ECG portable monitoring. This analysis, in addition to a single-electrode device proposed by \citet{lee-etal-2017-single-patch}, validates the feasibility and clinical viability of our full-set derivation from just two leads. Further analyses verify the necessity of DL in the reconstruction of ECG lead signals either by comparing DL performance to the previous interpolation approaches \cite{grande-fildago-etal-2021-ECG-proof, craven-etal-2017-ECG-proof} or by highlighting avenues facilitated only by the DL approach \cite{lee-etal-2017-single-patch}. Although these DL models accomplish their intended goals, they fail to classify multiple-disease targets much like the aforementioned structural CVD predictor rECHOmmend \cite{ulloa-cerna-2022-rECHOmmend} and functional CVD models \cite{zhou-etal-2021-functional-CVD}. Our approach aims to overcome the shortcomings of these models by using the conserved signal patterns identified by the GAN model to predict multiple-disease targets. \citet{hyo-chang-etal-2022-similar-one} attempted to reconstruct a full twelve-lead ECG from the slightly more ambitious one-lead input by also using a GAN model; however, their study is limited to reconstruction solely considering the limb lead I as the input and does not aim to classify CVD motifs in the reconstructed signals. Therefore, our work is distinguished because our model will take any single lead's signal to not only reconstruct the remaining signals but to identify highly conserved signal patterns and use the motifs to predict CVD onset: a more informative task. \\
\indent
The methods of our approach can be divided into two categories: pre-processing (\ref{ssec:pre-processing}) and model design (\ref{ssec:ganmodel}, \ref{ssec:cross-correlationanalysis}, \ref{ssec:cvd-predictivecnnmodel}). There has been a concerted effort to use DL to denoise ECG signals \cite{haroon-singh-2021-denoise, singh-pradhan-2021-denoising} and to classify the various components of the heartbeat \cite{burguera-2019-QRS-detection, he-etal-2018-heartbeat-classification}—both of which are used as inputs for our GAN model.  While these efforts have fruitful results, implementing these models in our pre-processing efforts may introduce training bias (or they may not be substantially cross-validated to be compatible with our dataset) which will be exasperated in the downstream components of our approach; therefore, we adapted a purely mathematical Daubechie wavelet denoising and a non-DL-based heartbeat classification pre-processing procedure validated by \citet{hazra-byun-2020-SynSigGAN} and \citet{kachuee-etal-2018-heartbeat-classification-mathematical} respectively. \\
\indent 
\citet{zhu-etal-2019-bigridlstm} was the first to show that Electrocardiogram (ECG) signals can be generated using a generative adversarial network (GAN) model where a bidirectional long short-term memory (LSTM) framework is used as a generator while a convolutional neural network (CNN) is used as the discriminator. The model was successfully used to synthesize complete 12-signal data sets from patient features. Given the great success of this framework, we believe that adapting it to take signal inputs instead of patient features can improve the quality of the signal generated allowing for effective classification. \citet{hazra-byun-2020-SynSigGAN} used the same model architecture as \citet{zhu-etal-2019-bigridlstm} and significantly improved the resolution of the signals generated by tuning the pre-processing and updating both the convolutional and pooling layers of the CNN discriminator. The GAN model's resulting signals will subsequently be classified (\ref{ssec:cross-correlationanalysis})—aided by this improved reconstructed signal resolution—providing further support for the architecture. The 15s twelve-lead ECG data used by our analysis is also the input format used in both the adapted pre-processing methods \cite{hazra-byun-2020-SynSigGAN, kachuee-etal-2018-heartbeat-classification-mathematical} and the various architectures \cite{zhu-etal-2019-bigridlstm, hazra-byun-2020-SynSigGAN}. \\
\indent  
\subsection{Mathematical Feature Extraction}
\label{ssec:mathematicalfeatureextraction}
\citet{ramli-ahmad-2003-cross-correlation} proved that by taking the cross-correlation between any two signals and by comparing the auto-correlation function of the signals to the original signal, the dependent features can be extracted. Attempts to verify this process have produced successful models identifying specific arrhythmia-indicating signal features \cite{chiu-etal-2005-correlation-arrhythmia} and even drowsiness predictors \cite{lee-etal-2017-drowsiness}. The analysis of \citet{ramli-ahmad-2003-cross-correlation}  was performed between leads from the same person; therefore, the novelty of our approach does not derive from the use of this technique, but because it is used to unveil the properties of the lead signals that the GAN model preserves. By piercing the black box this way, transfer learning can occur—i.e., applying the learning of ECG signal synthesis to the prediction of CVD. \\ 
\indent
\subsection{Deep Learning Methods for
Predicting Cardiovascular Disease}
\label{ssec:dlcvd}
CNN models have been remarkably successful using annotated ECG signal inputs to predict CVD of a signal type \cite{wu-etal-2021-CNN1, shankar-etal-2020-CNN2, dutta-etal-2020-CNN3}. \citet{sajja-kalluri-2020-CNN-superiority} validated the superiority of CNN models in this regard by comparing their performance to that of various other machine-learning algorithms. Furthermore, attempts to involve CNN models in CVD-predictive transfer learning have had preliminary successes and thus they can be the vehicle of multiple-disease prediction \cite{weimann-conrad-2021-CNN-transfer-learning}. Considering this, a predictive CNN model taking the conserved signal motifs identified in the signal cross-correlation feature extraction was implemented (\ref{ssec:cvd-predictivecnnmodel}). The patient functional outcome data used in our analysis compares favorably in terms of its comprehensiveness to the data used in the aforementioned analysis \cite{wu-etal-2021-CNN1, shankar-etal-2020-CNN2, dutta-etal-2020-CNN3}. While all three main components of our approach  have been explored independently, the use of each in succession to solve the next task's flaws produces our study's novelty.\\
\indent
Herein, we focus on predicting mutliple-CVD onset by presenting a bidirectional grid LSTM-CNN GAN model to reconstruct ECG signals (\ref{ssec:ganmodel}), identifying the features the GAN model conserves (\ref{ssec:cross-correlationanalysis}), and using the features as inputs for a predictive model (\ref{ssec:cvd-predictivecnnmodel}). 
\section{Approach}
\label{sec:approach} 
\begin{figure}[htbp!]
\centering
\includegraphics[width=\columnwidth]{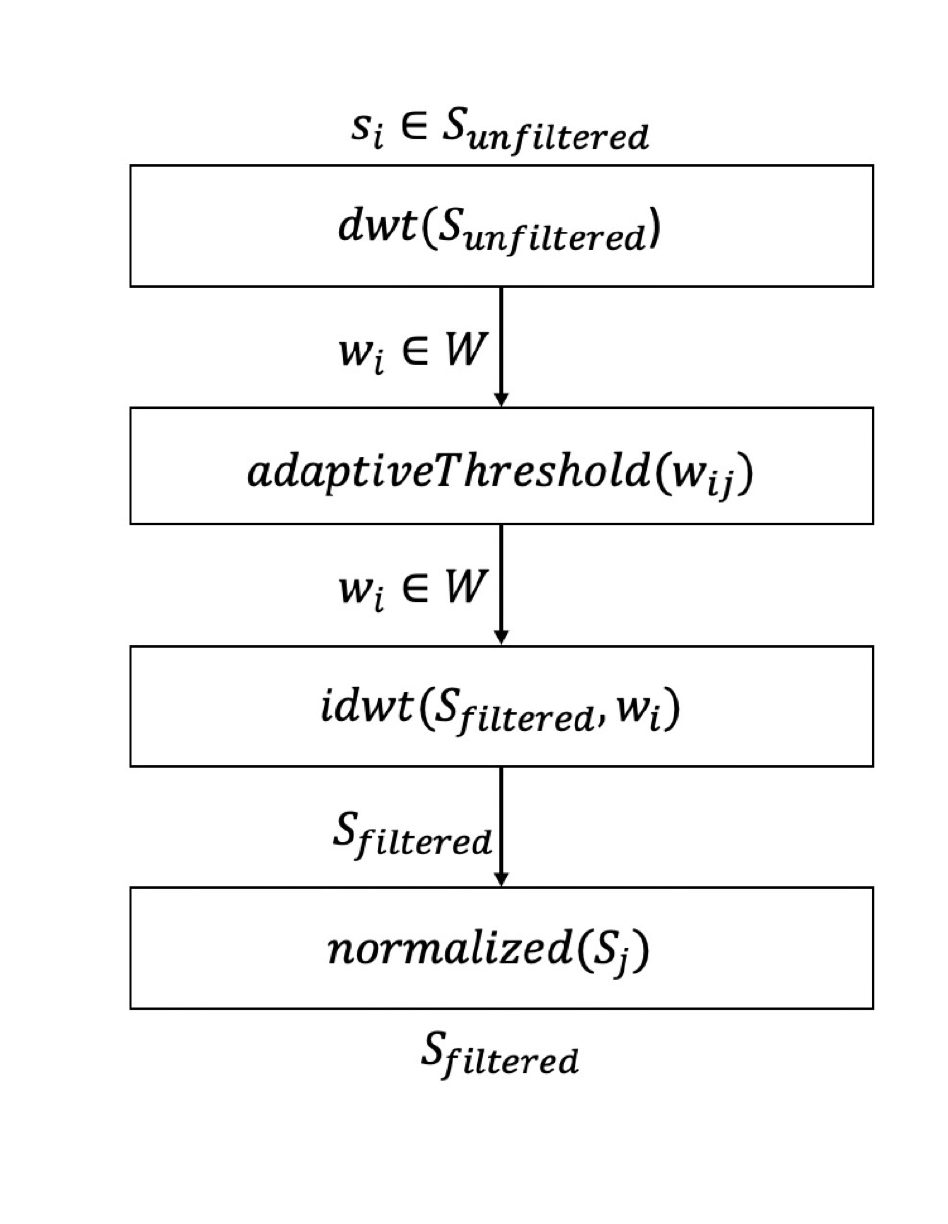}
\caption{Progression of pre-processing $P: S\textsubscript{unfiltered} \mapsto P: S\textsubscript{filtered}$.}
\label{fig:Fig1}
\end{figure}

Here, the pre-processing procedure (Figure \ref{fig:Fig1}) that the individual 15s ECG lead signals are subjected to is outlined. 
The procedure (\ref{ssec:pre-processing}) outputs the denoised signal. The architecture of the Generative Adversarial Network (GAN) model (\ref{ssec:ganmodel}) that inputs the pre-processed signal is subsequently detailed along with the signal-correlation analysis (\ref{ssec:cross-correlationanalysis}) performed on the model's outputs. Then, the cardiovascular disease (CVD)-predictive model (\ref{ssec:cvd-predictivecnnmodel}) that inputs the features identified by the signal-correlation analysis is described. Overall, the approach facilitates the evolution of  an input of any 15s lead signal to the output of a CVD prediction. The approach is executed in Python 3.10 \cite{van-drake-1995-python}. 
\subsection{Pre-processing}
\label{ssec:pre-processing}
Pre-processing takes each 15s electrocardiogram (ECG) lead signal as input S\textsubscript{unfiltered} and denoises them (Algorithm \ref{alg:denoise}). There are n\textsubscript{filters} = 4 high-low pass Daubechie discrete wavelet transforms (\textit{dwtHP}  and \textit{dwtLP} respectively) from the pywt 1.4.1 package \cite{gregory-etal-2022-pywt} performed. Upon completion of each \textit{dwt\textsubscript{i}}, the low pass signal \textit{cA\textsubscript{i}} is subsampled (\textit{resample}) by two and passed to the next round of filters while the high pass signal \textit{cD\textsubscript{i}} is subsampled by two (\textit{resample}) and extracted to become a wavelet (\textit{w\textsubscript{i}}). After all filters are applied, adaptive thresholding is performed on each of the \textit{i} wavelets at each signal time point \textit{j} with moving average window of length \textit{r} = 32 \cite{hazra-byun-2020-SynSigGAN}: \\
\begin{equation}
    \begin{aligned}
        adaptiveThreshold(w_{ij}) = \\1/r\sum_{n=r*(j-1)}^{rj-1}\mid w_{1j}\mid\ *2^{i}
  \end{aligned}
\end{equation}
To construct the denoised signal, the inverse Daubechie DWT (\textit{idwt}) function from the pywt 1.4.1 package \cite{gregory-etal-2022-pywt} is performed on the threshold-ed wavelets (\textit{w\textsubscript{i}}). These denoised signals (\textit{S}) at each time point \textit{j} are subsequently normalized to be between zero and one: \\
\begin{equation}
    \begin{aligned}
normalized(S_j) = (S_j - min(S))/\\(max(S)-min(S))
  \end{aligned}
\end{equation}
Examples of the pre-processing output are provided in (\ref{ssec:pre-processingvalidation}).
\begin{algorithm}[htb!]
\caption{Daubechie denoising algorithm} \label{alg:denoise}
\textbf{Input:} S\textsubscript{unfiltered}, n\textsubscript{filters} \\
\textbf{Output:} S\textsubscript{filtered}
\begin{algorithmic}[1]
\State $cA_0 \gets resample(dwtLP(S\textsubscript{unfiltered}))$
\State $cD_0 \gets dwtHP(S\textsubscript{unfiltered})$
\State $w_0 \gets resample(cD)$
\For{i=2 to n-1}
    \State $cA_i \gets resample(dwtLP(cA\textsubscript{i-1}))$
    \State $cD_i \gets dwtHP(cA\textsubscript{i-1})$
    \State $w_i \gets resample(cD_i)$
\EndFor
\State $w_n \gets resample(cA\textsubscript{n-1})$
\For{i=2 to n}
    \State $w_i \gets adaptiveThreshold(w_i)$
\EndFor
\State $S\textsubscript{filtered} \gets idwt(w_n,w\textsubscript{n-1})$
\For{i=n-2 down to 1}
    \State $S\textsubscript{filtered} \gets idwt(S\textsubscript{filtered},w\textsubscript{i})$
\EndFor
\end{algorithmic}
\end{algorithm}

\subsection{GAN Model}
\label{ssec:ganmodel}
\begin{figure}[htbp!]
\includegraphics[width=\columnwidth]{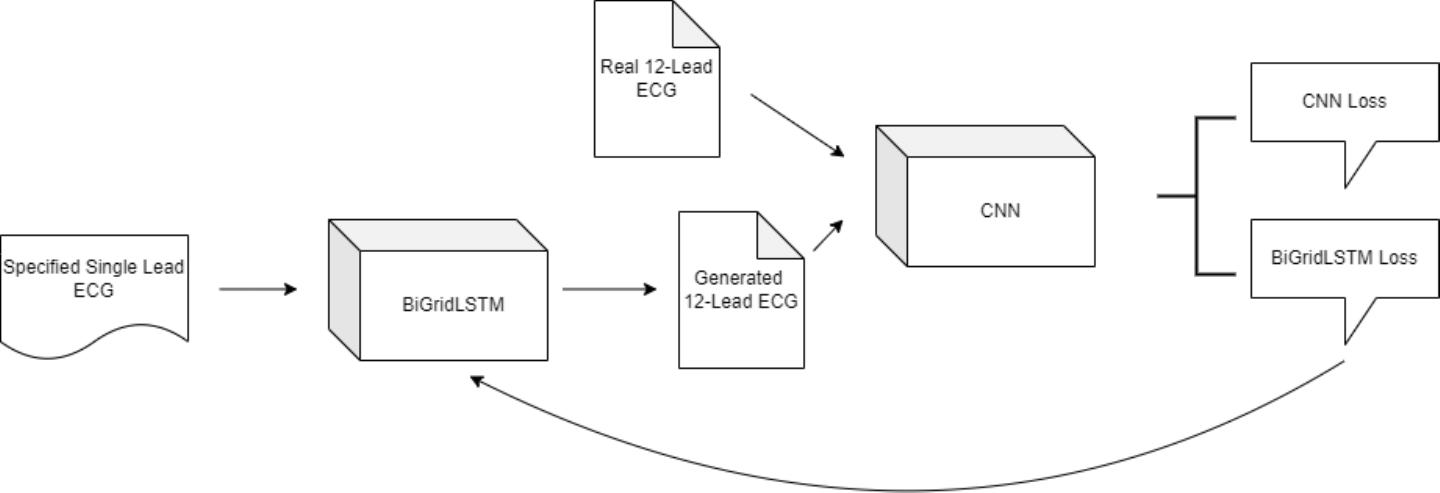}
\centering
\caption{Overview of the GAN model. The generator is a bidirectional LSTM that takes a single lead input and the discriminator is a CNN that attempts to distinguish the real signal from the generated signal.}
\label{fig:GANOverview}
\end{figure}

Our GAN model consists of a bidirectional long short-term memory (LSTM) generator that aims to recreate ECG lead signals—from a two-signal input—that the discriminator cannot distinguish from the patient's actual signal (Figure \ref{fig:GANOverview}). 

The discriminator of the GAN model is a convolutional neural network (CNN) that attempts to differentiate the artificial signals from the patients' actual lead signals. Mathematically, the baseline GAN model can be modeled as a minmax game represented by the generator loss function—i.e., the penalties paid by either the generator or discriminator for failing to perform their respective tasks. The generator aims to minimize this loss while the discriminator attempts to maximize the loss for the generator. To mathematically represent the situation, the domain of all \textit{N} pre-processed signals can be defined as ${s_i} \in S\textsubscript{filtered}$ while the \textit{M} signals of the generator output, or in other words, the discriminator domain, are defined as ${q_i} \in Q$. The generator is thus a mapping defined as $G: S\textsubscript{filtered} \mapsto Q$. The loss function can then be defined as
\begin{equation} \label{eq:3}
  \begin{aligned}
    \mathcal{L}(G, D, S\textsubscript{filtered}, Q)= \\
    min_{D}max_{G}\: \mathbb{E}_{q \sim p(q)}[log(D(q)]   \\
    + \mathbb{E}_{s \sim p(s)}[log(1-D(G(s)))] 
  \end{aligned}
\end{equation}
where \textit{D} is the discriminator, $\mathbb{E}_{q \sim p(q)}$ is the expectation of the probability given by the discriminator of a real signal being real on random input $p(q)$ over the distribution of ${q}$, and $\mathbb{E}_{s \sim p(s)}$ is the expectation of the probability given by the discriminator of a generated signal being real on random input $p(s)$ over the distribution of ${s}$ \cite{Zhu-etal-2017-loss-function}. \\
\indent 
\subsubsection{Bidirectional LSTM Generator}
\label{sssec:generator}
The baseline LSTM structure allows for long term dependencies to be considered in the model architecture. A LSTM is generally defined in terms of cell states which control what information is considered by the model through a series of gates. The sigmoid forget layer removes information from consideration, the sigmoidal input gate selects values of the cell state to be updated, the input hyperbolic tangent gate inputs new candidate values to the cell state, and finally sigmoidal and hyperbolic tangent output layers select the output information \cite{hochreiter-schmidhuber-1997-lstm}. The cell state output is subsequently passed to the corresponding layer of the underlying recurrent neural network (RNN)—a type of neural network where the hidden layer outputs can be used to update prior layers—preserving the long term dependencies \cite{Sherstinsky-2020-RNN}. This process is susceptible to the vanishing gradient problem: the gradient of the backpropagation updating the sigmoidal layers approaches zero due to the small derivatives at the sigmoid function extremes. The vanishing gradient problem prevents further model training. To prevent the gradient from vanishing, the baseline LSTM structure can be extended into a bidirectional grid. Instead of the information being passed from the hidden layers only temporally, the layers and corresponding cell states are stacked on top of each other to add a depth layer that decreases the probability that the backpropagation produces a near-zero gradient \cite{schuster-paliwal-1997-bigrid}. This more advanced, bidirectional grid approach is implemented as the generator $G: S\textsubscript{filtered} \mapsto Q$ according to the LSTM state variables, time block, and depth block outlined by \citet{hazra-byun-2020-SynSigGAN}. The approach of \citet{hazra-byun-2020-SynSigGAN} is favored to the bidirectional LSTM-based ECG signal generator presented by \citet{zhu-etal-2019-bigridlstm} because of the explicit consideration of the vanishing gradient by \citet{hazra-byun-2020-SynSigGAN}. \\
\indent
\subsubsection{CNN Discriminator}
\label{sssec:discriminator}
The time-series nature of the ECG signal confers the use of one-dimensional convolutions in the CNN discriminator as opposed to the more traditional two-dimensional convolution used in computer vision \cite{zhu-etal-2019-bigridlstm}. A one-dimensional CNN is modeled in terms the number and size (input and output) of its convolution and pooling layers, number of kernels per layer, and final activation function \cite{serkan-etal-2021-1dCNN} as defined for the discriminator here (Table \ref{tab:table2CNN}). The kernels in each convolutional layer serve as feature filters that parse through both the generated and real ECG signals according to the defined stride length—computing the dot product between the kernel and signal while reducing the dimensions of the data \cite{kiranyaz-etal-2019-1dCnn}.  Each kernel produces a feature map that represent certain characteristics of the ECG signal. To prevent over-fitting, the pooling layers sub-sample the feature maps. The final convolution produces a fully connected hidden layer whose outputs are subjected to the softmax activation function which gives the probability that the signal is generated or real \cite{iwana-etal-2019-softmax}. The probabilities are passed to the aforementioned loss function (Equation \ref{eq:3}).

\subsection{Cross-correlation analysis}
The baseline signal correlation test is the cross-correlation function computed as
\begin{equation}
r(q,s)=  \sum_{j=0}^{T} q_{j}s_{j}
\end{equation}
where $q \in Q$ and $s \in S_{filtered}$ for all time points $T$ in the ECG signals \cite{podobnik-stanley-2008-cross-correlation}. For each of the 12 ECG leads, the cross-correlation function between every synthesized lead signal and the original patient signal is calculated and averaged across the entire data set. The function is subsequently charted and analyzed by cardiologists at the Robert Wood Johnson University Hospital for feature extraction. The efficiency of the feature extraction is compared to the studies by \citet{ramli-ahmad-2003-cross-correlation}, \cite{chiu-etal-2005-correlation-arrhythmia}, and \cite{lee-etal-2017-drowsiness}.
While the cross-correlation function acts as a baseline mechanism for extracting the latent relationships learned by the GAN model, the auto-correlation function defined
\begin{equation}
r(q, i)=  \sum_{j=i}^{T} q_{j}q_{j-i}
\end{equation}
where $i$ is a time-translation introduced to the generated signal, may provide further, more definitive features \cite{ramli-ahmad-2003-cross-correlation}. This is because repeated, important patterns from each heartbeat should be perpetuated throughout the entirety of the signal which would only be captured by the auto-correlation function. The auto-correlation function is once again charted and analyzed by cardiologists at the Robert Wood Johnson University Hospital for feature extraction. The analysis done by \citet{ramli-ahmad-2003-cross-correlation} serves as the baseline comparison. 
\label{ssec:cross-correlationanalysis}
\subsection{CVD-Predictive CNN Model}
\label{ssec:cvd-predictivecnnmodel}
We cannot write the approach for this section yet because it depends on how successful the aforementioned cross-correlation analysis is. Until we know what features are successfully identified, we cannot design a model structure; however, we include it here because the novelty of our research is not apparent without this model. We will implement it in the future. We anticipate using a CNN-predictive model as the models that have come closest to achieving multiple-CVD onset predictions are CNN-based models. That being said, the novelty of the analysis derives from utilizing all three components: using the GAN model to  understand the latent relations of ECG signals, the cross-correlation analysis to identify the latent connections the GAN model is making, and the CVD-predictive model to predict multiple-CVD onsets which has never been done before successfully. The approach is generalizable as it can take any ECG input to train the GAN model, update the cross-correlation analysis, and fine-tune the CVD-predictive model. Furthermore, the ultimate aim of the approach is so that the resulting CVD-predictive model is generalizable for any ECG signal input and most CVD predictions. 

\section{Experiments}
\label{sec:experiments}
We implement our models in TensorFlow and experiment with three different generative models: our baseline GAN model, the one-lead predictive model by \cite{hyo-chang-etal-2022-similar-one}, and our advanced bidirectional LSTM-1d CNN GAN model. All our models are evaluated on our dataset (\ref{ssec:myovista}). 
\subsection{MyoVista Dataset}
\label{ssec:myovista}
Our MyoVista dataset consists of 15s 12-lead ECG data measured using MyoVista's wavECG patient data from three different hospitals: West Virginia University, Mount Sanai, and Windsor. Functional outcome data for each patient in the 1000-patient dataset is recorded. Patient ages ranged from 18-96 years old with a mean age of $56.8\pm{15}$ and patient gender is 46.6\% and 53.4\% female and male respectively. Each race is represented in the data set, but Caucasians are overrepresented (Table \ref{tab:racetable}). Weight measurements (mean $86.8\pm{23}$kg) indicate a well-distributed sample with SD accounting for more than $25\%$ of the mean. The distribution of height (mean $169.5\pm{10}$cm) is more conservative but indicative of a distributed sample. Over 2000 functional outcomes are recorded—e.g., ejection fraction, Rsign, and blood pressure—for each patient. This dataset was collected because with a large array of functional outcomes, we expect to be able to extract features from the generated signals that can then be used to predict at least some of the functional outcomes. Furthermore, with MyoVista ECG signal data having been used to successfully predict CVD-onset in the past \cite{sengupta-etal-myovista}, signal data from the wavECG is used. Due to HIPAA regulations, other datasets cannot be accessed and therefore, the models are only evaluated on our dataset. 

\begin{table}[htbp!]
\centering\resizebox{\columnwidth}{!}{
\begin{tabular}{c|cccccc} 
\toprule
\bf  & \bf Caucasian & \bf African-American & \bf Hispanic & \bf Asian & \bf Mean \\
\midrule
\% & \bf 80.5 & 5.7 & 3.7 & 6.3 & NA\\
\% Male & 49.7 & 37.1 & 29.3 & \bf 58.5 & 53.4 \\
Age (years) & $\bf56.7\pm{17}$ & $53.1\pm{13}$ & $53.4\pm{14}$ & $50.6\pm{15}$ & $56.8\pm{15}$ \\
Weight (kg) & $89.2\pm{25}$ & $\bf93.2\pm{26}$ & $79.8\pm{17}$ & $68.1\pm{14}$ & $86.8\pm{23}$ \\
Height (cm) & $\bf169.6\pm{10}$ & $169.3\pm{10}$ & $164.8\pm{8.7}$ & $164.8\pm{8.4}$ & $169.5\pm{10}$ \\
\bottomrule
\end{tabular}}
\caption{Percentage of MyoVista dataset represented by each race along with the racial breakdown of the generic patient data. The largest values for each category are bolded.}
\label{tab:racetable}
\end{table}

\subsection{Pre-processing Validation}
\label{ssec:pre-processingvalidation}
\begin{figure}[htbp!]
\centering

\begin{subfigure}{\columnwidth}
\centering
\includegraphics[width=\columnwidth,height=17mm]{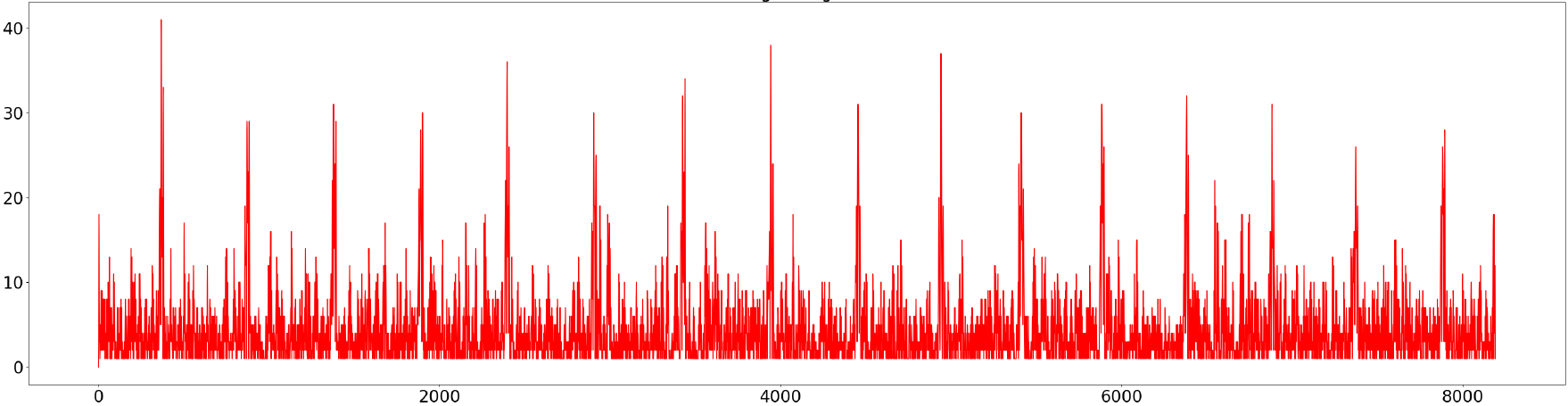}
\caption{Sub-figure 1. The original signal recorded by wavECG.}
\label{tab:original-1}
\end{subfigure}

\begin{subfigure}{\columnwidth}
\centering
\includegraphics[width=\columnwidth,height=17mm]{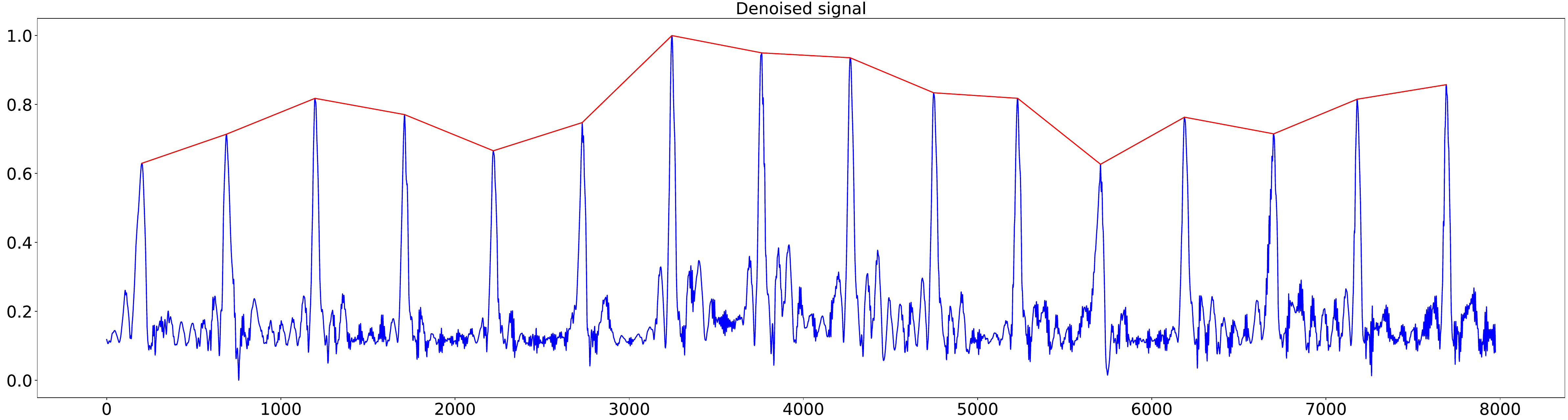}
\caption{Sub-figure 2. A representative denoised signal. The blue line is the result of the original signal being subjected to four high-low pass Daubechie discrete wavelet transforms, adaptive thresholding, four high-low pass Daubechie inverse wavelet transforms, and normalization as described in Section \ref{ssec:pre-processing}. The red line serves as proof that the algorithm is able to successfully identify each peak—i.e. R complex—of each heartbeat.}
\label{tab:denoised-1}
\end{subfigure}

\caption{Comparison of the original signal recorded by wavECG and the denoised signal after the procedure outlined in Section \ref{ssec:pre-processing} is performed.}
\label{tab:pre-processing-fig}
\end{figure}

For each signal recorded by the wavECG, the pre-processing procedure outlined in Section \ref{ssec:pre-processing} is performed. The denoised signals clearly display the correct number of heartbeats and fundamental components of a heartbeat: the P-Q segment, the QRS complex, and the S-T segment (Figure \ref{tab:denoised-1}). Furthermore, for each R-peak in the original signal (Figure \ref{tab:original-1}), there is an R-peak in the denoised signal (Figure \ref{tab:denoised-1}), further validating the correctness of the pre-processing procedure. Upon completion of the pre-processing procedure, the R-peak time intervals are calculated and each heartbeat is segmented. Padding is added to the signal according to the maximum R-peak time interval across all patients.

\begin{table}[htbp!]
\centering\resizebox{\columnwidth}{!}{
\begin{tabular}{c|ccc} 
\toprule
\bf Set & \bf Total Number of Heartbeats & \bf Heartbeats per Lead \\
\midrule
Training & 1,584,000 & 132,000\\
Validation & 198,000 & 16,500\\
Test & 198,000 & 16,500\\
\bottomrule
\end{tabular}}
\caption{Data split for the GAN model inputs (Section \ref{ssec:generationofecgsignals})}
\label{tab:trainvalidationtest}
\end{table}

For the GAN model (Section \ref{ssec:generationofecgsignals}), the sequential time-series representation of each heartbeat constitutes the input. For each of the 1000 patients, there are 15 heartbeats. For each of the patient's lead's heartbeats, we are trying to predict the other 11 heartbeats; therefore, there are $1000*15*12*11=1,980,000$ model inputs. Because of the size of the pre-processed dataset, we randomly shuffled the heartbeats and allocated 80\% for training, 10\% for validation, and 10\% for training (Table \ref{tab:trainvalidationtest}).

\subsection{Generation of ECG signals}
\label{ssec:generationofecgsignals}
\begin{figure}[htbp!]
\centering
\includegraphics[width=\columnwidth]{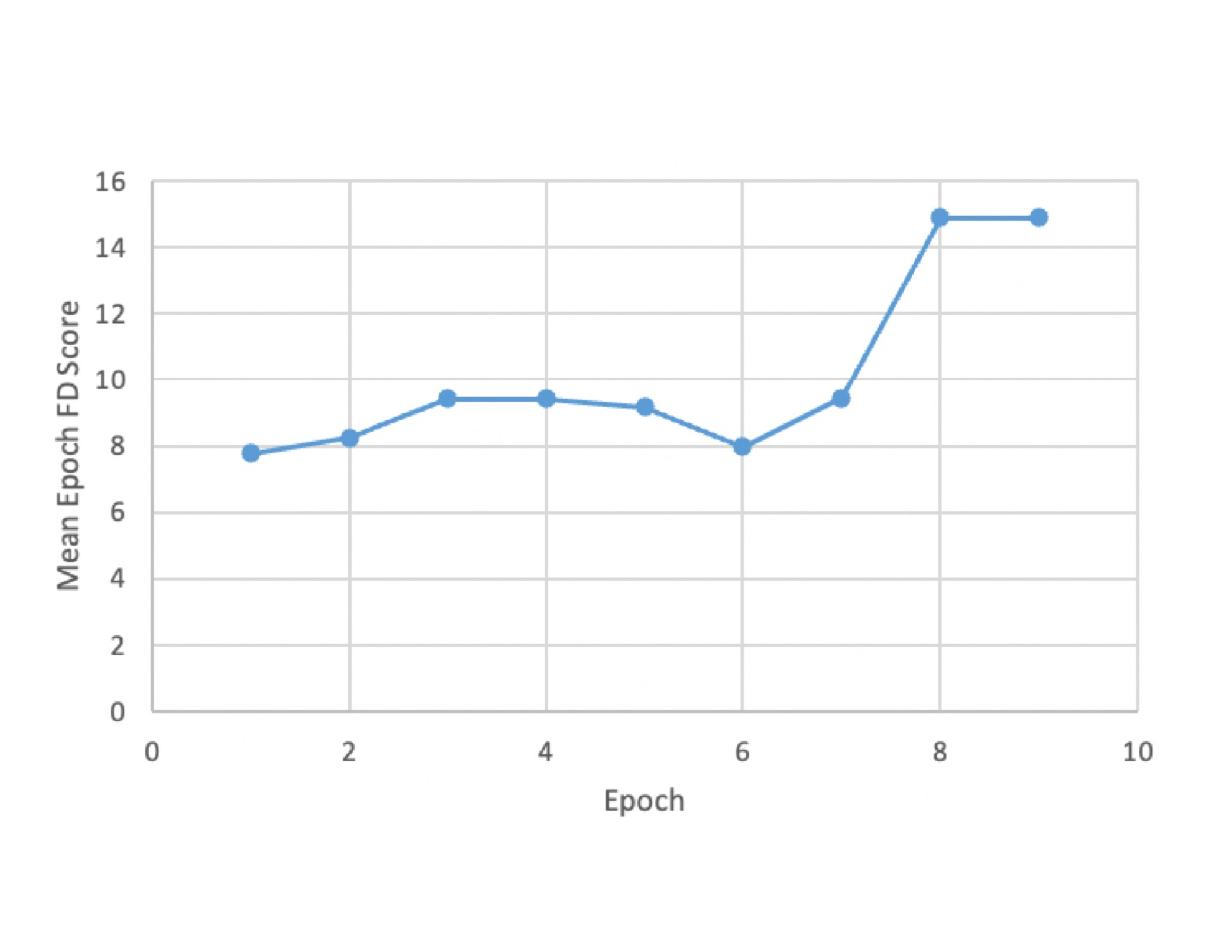}
\caption{The model's mean FD score on the validation score after each epoch.}
\label{fig:epoch}
\end{figure}

To reiterate, we train a basic  LSTM generator and one-layer 1d-CNN discriminator for our baseline GAN model. We then compare the results to our advanced GAN model with the bidirectional LSTM generator and multi-layer 1d-CNN discriminator optimized for each lead. Both of these results are subsequently compared to \citet{hyo-chang-etal-2022-similar-one} GAN model that was specifically designed to synthesize the signals from just one lead. Note we did not have the requisite time to the model from the \citet{hyo-chang-etal-2022-similar-one} study, but instead only present it here as a data point. In the future, we will reconstruct the model and test it on our dataset and compare the results to our model.\\
\indent
To evaluate how similar the generated signal is to the input signal, we use Fréchet Distance (FD) defined as:
\begin{equation}
  \begin{aligned}
    FD(s \in S_{filtered}, q \in Q) = \\
    min(max_{i=1,...,n}(||s_i, q_i||))
  \end{aligned}
\end{equation}
where $||s_i, q_i||$ is the Euclidean distance between points $s_i$ and $q_i$. In other words, FD computes the maximum distance of $s_i$ and $q_i$ in a given alignment for all alignments of \textit{n} points from \textit{s} and from \textit{q} and then finds the minimum of these maximum distances. Therefore, FD evaluates similarity based on both the ordering and location of the signals' points. Both factors are considered in the cross-correlation analysis (\ref{ssec:cross-correlationanalysis}), thus FD is the chosen similarity metric. Because the model proposed by \citet{hyo-chang-etal-2022-similar-one} is specifically designed for the limb lead I as the sole input, the individual FD scores for this model will likely not be as low, but comparable—i.e., within one standard deviation. \\
\indent
Our baseline model is trained for nine epochs using the learning rate of $1\mathrm{e}{-4}$. Note that because we did not have enough time to optimize hyperparameters, we did not optimize the learning rate but we did optimize the number of epochs. We saved the model after each epoch so that the epoch that produced the lowest FD during validation could be selected for the generation. Epoch 1 was selected because it produced the lowest FD score in the validation (Figure \ref{fig:epoch}).

The input batch size was 256. Cross-entropy was used as the loss function and the Adam algorithm was used as our optimizer. For the bidirectional LSTM generator, a time-distributed 1-unit dense layer with a sigmoidal activation was applied. The details of the CNN discriminator are presented in Table \ref{tab:table2CNN}. The Relu activation function is used for the convolution and final convolution (FC) layers.

\begin{table}[htbp!]
\centering\resizebox{\columnwidth}{!}{
\begin{tabular}{c|ccc} 
\toprule
\bf Layer & \bf Number of Kernels & \bf Filter Size & \bf Stride\\
\midrule
$C_1$ & 64 & 3 & 3\\
$P_1$ & 1 & 2 & 2\\ 
$C_{...}$ & 64 & 3 & 3\\
$P_{...}$ & 1 & 2 & 2\\
FC & 100 & 1 & 1\\
Softmax & 1 & 1 & 1\\
\bottomrule
\end{tabular}}
\caption{The definition of each CNN layer of the GAN discriminator}
\label{tab:table2CNN}
\end{table}

\begin{figure}[htbp!]
\centering
\includegraphics[width=\columnwidth]{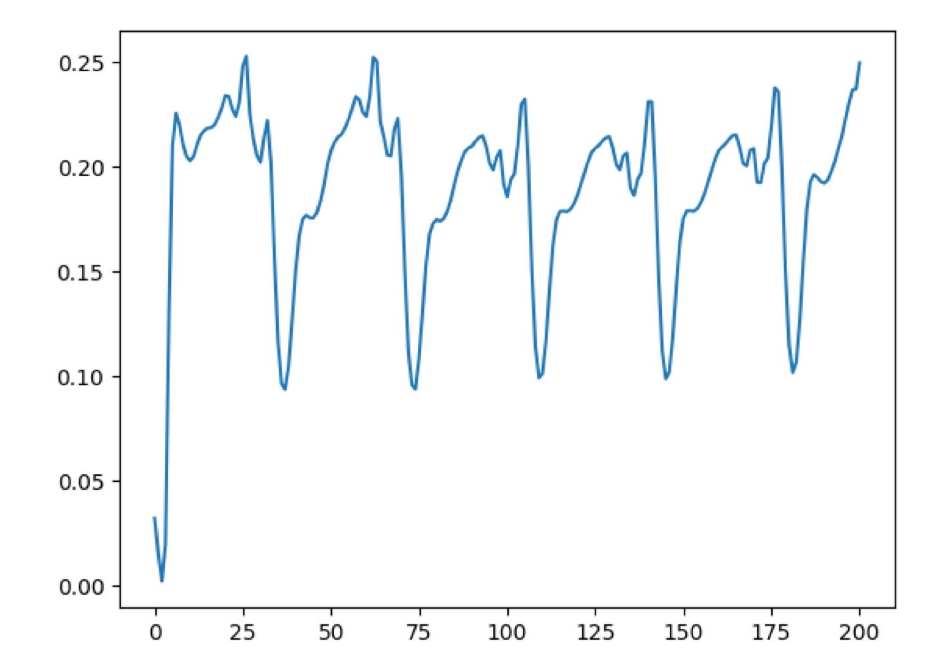}
\caption{Representative generated signal by the baseline model. Each heartbeat of the signal clearly contains the P-Q segment, the QRS complex, and the S-T segment.}
\label{fig:generatedsignal}
\end{figure}

The mean FD scores comparing our baseline model and the one-lead model created by \citet{hyo-chang-etal-2022-similar-one} are presented in Table \ref{tab:results}. Our baseline model outperforms the model by \citet{hyo-chang-etal-2022-similar-one} in two of the three experiments. Further inspection of the experiments by \citet{hyo-chang-etal-2022-similar-one} indicates FD scores for the limb leads that are far lower than ours but far higher FD scores for the pre-cordial leads. Biologically, this can be explained by visualizing our prediction (Figure \ref{fig:generatedsignal}). Our generated signal clearly preserves all of the fundamental components of a heartbeat: the P-Q segment, the QRS complex, and the S-T segment. These fundamental components of the heartbeat are better preserved in the pre-cordial leads as they are closer to the heart; therefore, it appears that our model is trained better to generate the pre-cordial signals than it is to the limb leads. This could be explained by the noisiness of the limb leads prior to pre-processing. The limb leads were noticeably more noisy before the pre-processing and although the signals were processed successfully, the QPRST components were less distinct for the limb leads than the pre-cordial leads.

\begin{table}[htbp!]
\centering\resizebox{\columnwidth}{!}{
\begin{tabular}{c|ccccc} 
\toprule
\bf Metric & \bf Baseline & \bf \citet{hyo-chang-etal-2022-similar-one} E1 & \bf \citet{hyo-chang-etal-2022-similar-one} E2 & \bf \citet{hyo-chang-etal-2022-similar-one} E3\\
\midrule
FD & 7.77 & 9.062 & 8.124 & 6.071\\
\bottomrule
\end{tabular}}
\caption{Mean FD scores from our baseline model, our advanced model, and the three experiments that \citet{hyo-chang-etal-2022-similar-one} used.}
\label{tab:results}
\end{table}

\subsection{Features of ECG Signals that Predict CVD}
\label{ssec:featuresofecgsignalsthatpredictcvd}
\begin{table}[htbp!]
\centering

\begin{subtable}{\columnwidth}
\centering\small{ 
\begin{tabular}{c|ccc} 
\toprule
\bf & \bf 1 & \bf 2 & \bf 3 \\
\midrule
1 & N/A & 0.53 & -0.33 \\
2 & 0.53 & N/A & 0.48 \\
3 & -0.33 & 0.48 & N/A \\
4 & -0.83 & -0.88 & -0.11 \\
5 & 0.80 & 0.05 & -0.77 \\
6 & 0.14 & 0.86 & 0.82 \\
7 & -0.38 & -0.35 & -0.01 \\
8 & 0.02 & -0.08 & -0.15 \\
9 & 0.30 & 0.29 & 0.02 \\
10 & 0.45 & 0.46 & 0.06 \\
11 & 0.55 & 0.52 & 0.01 \\
12 & 0.57 & 0.53 & 0.01 \\
\bottomrule
\end{tabular}}
\caption{Sub-table 1. Precordial leads 1-3}
\label{tab:Table 1-1}
\end{subtable}
\vspace{0.5em}

\begin{subtable}{\columnwidth}
\centering\small{ 
\begin{tabular}{c|ccc} 
\toprule
\bf & \bf 4 & \bf 5 & \bf 6 \\
\midrule
1 & -0.83 & 0.80 & 0.14 \\
2 & -0.88 & 0.05 & 0.86 \\
3 & -0.11 & -0.77 & 0.82 \\
4 & N/A & -0.42 & -0.56 \\
5 & -0.42 & N/A & -0.35 \\
6 & -0.56 & -0.35 & N/A \\
7 & 0.43 & -0.18 & -0.19 \\
8 & 0.06 & 0.15 & -0.10 \\
9 & -0.30 & 0.22 & 0.22 \\
10 & -0.49 & 0.28 & 0.33 \\
11 & -0.58 & 0.35 & 0.33 \\
12 & -0.60 & 0.37 & 0.34 \\
\bottomrule
\end{tabular}}
\caption{Sub-table 2. Precordial leads 4-6}
\label{tab:Table 1-2}
\end{subtable}

\caption{The average correlation coefficient between each of the first three precordial leads signals (Table \ref{tab:Table 1-1}) and the other eleven leads. The average correlation coefficient between each of the other precordial leads signals of IV, V, and VI (Table \ref{tab:Table 1-2}) and the other eleven leads. All patient data were included in the calculations.}
\label{tab:Table 1}
\end{table}
To determine baseline relationships between the more informative precordial leads—i.e., the leads positioned next to the heart—and the other leads, the average correlation coefficient between each precordial lead and the other remaining eleven leads is calculated with consideration of all patients in the study prior to pre-processing (Table \ref{tab:Table 1}). The correlation coefficients taken here initially will eventually then be compared to the correlation coefficient results of the generated GAN model signals. Note that because this was a part of Max's individual project and we barely had enough time to finish the group experiment during this time (we are using our group member's GPU and we could not get time to perform the analysis) we could not repeat the analysis for the generated signals. This section is included once again to address the novelty of the project and will be completed for the final publication.

\subsection{Validation of Multiple-CVD Predictions}
This section is included once again to address the novelty of the project and will be completed for the final publication.
\label{ssec:validationofmultiple-cvdpredictions}


\section{Analysis}
\label{sec:analysis}
\begin{figure}[htbp!]
\centering
\includegraphics[width=\columnwidth]{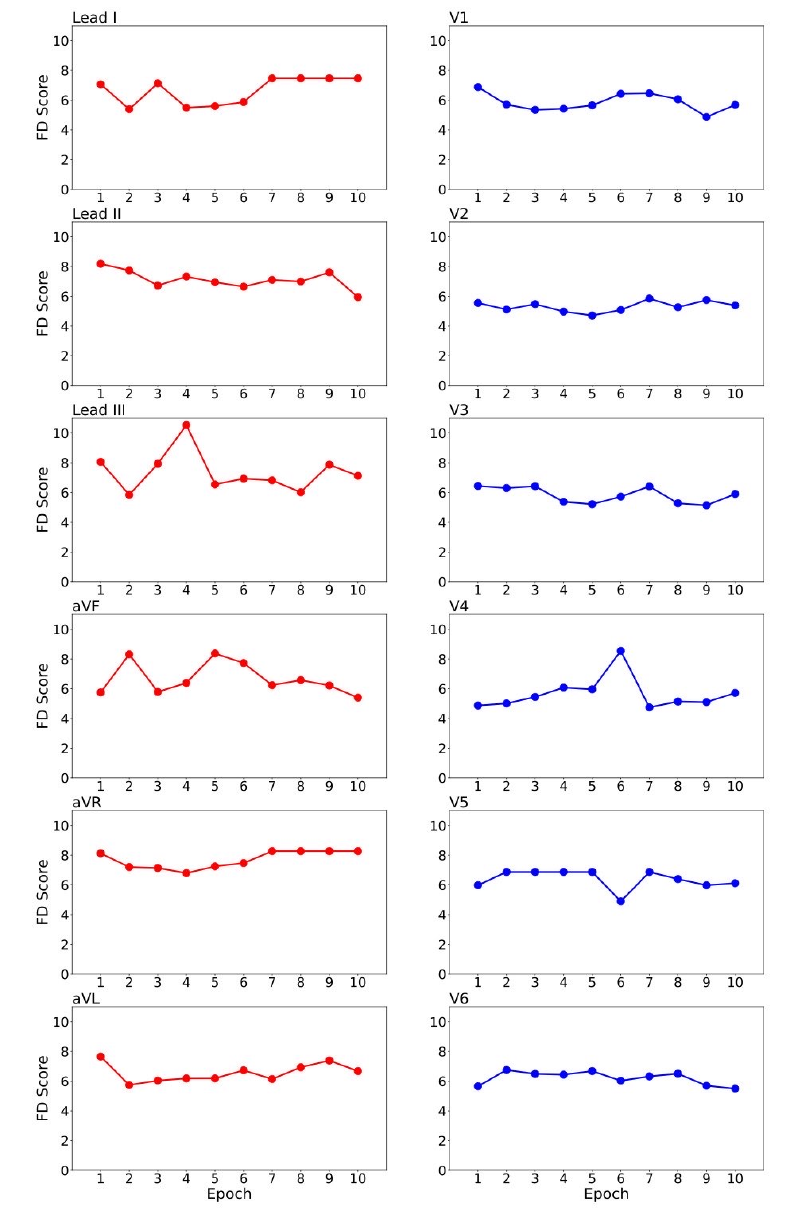}
\caption{The model's mean FD score on the validation set for the generated signal of each lead after each of the ten epochs. The mean FD scores for six limb leads (red) are on the left and the six pre-cordial leads (blue) are on the right. The best epoch for each lead is subsequently chosen (Table \ref{tab:analysistable}).}
\label{fig:leadepoch}
\end{figure}

In order to understand the biological underpinnings of the model's performance and to assess the model's performance in predicting the signals of specific leads, we split the validation and testing sets in terms of the predicted lead to produce the advanced model. In other words, we train 12 generators (and consequently 12 discriminators) which are each specialized in generating a particular lead. For each predicted lead, the epoch that performed best on the development set corresponding to that lead is used on the test set for that lead (Figure \ref{fig:leadepoch}). The batch size for the advanced model is 128. 

\subsection{Performance Analysis}
\label{ssec:performanceanalysis}
In Table \ref{tab:analysistable}, the mean FD score of each lead's best model's performance on the testing set is compared to the best mean FD score of \citet{hyo-chang-etal-2022-similar-one}. While our model's overall mean FD score of $6.38\pm{0.1}$ outperforms all three of \cite{hyo-chang-etal-2022-similar-one} models' mean FD scores, further inspection of the specific leads in which our model performs better yields a distinct trend. For each of the limb leads (Table \ref{tab:analysisepochlimb}), \citet{hyo-chang-etal-2022-similar-one} outperforms our model. Specifically, the signals generated for aVR and aVL by their model are much more similar to the patient's actual signal than our model; however, when examining the pre-cordial leads (Table \ref{tab:analysisepochprecordial}), our model consistently outperforms for every lead. When comparing within our model, only one of the two limb leads (AVR and AVF) better the worst performing pre-cordial lead (V6), and neither outperforms any of the other pre-cordial leads. This differentiation in model performance between the limb and pre-cordial leads perhaps can be explained by three different explanations: biological, mechanical, and computational. 

\begin{table}[htbp!]
\centering

\begin{subtable}{\columnwidth}
\centering\resizebox{\columnwidth}{!}{
\begin{tabular}{c|cccccc} 
\toprule
\bf & \bf I & \bf II & \bf III & \bf aVR & \bf aVL & \bf aVF\\
\midrule
Epoch & 1 & 9 & 1 & 9 & 3 & 1\\
Mean FD & \bf 5.76 & 6.28 & 5.74 & 5.45 & 7.06 & 5.50\\
\citet{hyo-chang-etal-2022-similar-one} FD & NA & \bf 4.55 & \bf 4.93 & \bf 0.94 & \bf 1.48 & \bf 5.25\\
\bottomrule
\end{tabular}}
\caption{Sub-table 1. Limb leads}
\label{tab:analysisepochlimb}
\end{subtable}
\vspace{0.5em}

\begin{subtable}{\columnwidth}
\centering\resizebox{\columnwidth}{!}{
\begin{tabular}{c|cccccc} 
\toprule
\bf & \bf V1 & \bf V2 & \bf V3 & \bf V4 & \bf V5 & \bf V6\\
\midrule
Epoch & 8 & 4 & 8 & 6 & 5 & 9\\
Mean FD & \bf 4.73 & \bf 4.89 & \bf 5.18 & \bf 4.77 & \bf 4.71 & \bf 5.55\\
\citet{hyo-chang-etal-2022-similar-one} FD & 13.64 & 10.85 & 17.64 & 15.27 & 13.60 & 11.50\\
\bottomrule
\end{tabular}}
\caption{Sub-table 2. Pre-cordial leads}
\label{tab:analysisepochprecordial}
\end{subtable}

\caption{The epoch producing the best FD score of each lead that is subsequently used in the prediction of that lead's signal is presented. The mean FDs from the model's prediction of each lead are then calculated and compared to the mean FD from \citet{hyo-chang-etal-2022-similar-one} for each lead. Lead I is omitted from \citet{hyo-chang-etal-2022-similar-one} because their model is designed to take lead I as input. The better mean FD of the two models is bolded.}
\label{tab:analysistable}
\end{table}

\begin{figure}[htbp!]
\centering

\begin{subfigure}{\columnwidth}
\centering
\includegraphics[width=\columnwidth]{tex/Figure1.pdf}
\caption{Sub-figure 1. Denoised limb lead representative sample}
\label{tab:denoised-limb}
\end{subfigure}

\begin{subfigure}{\columnwidth}
\centering
\includegraphics[width=\columnwidth]{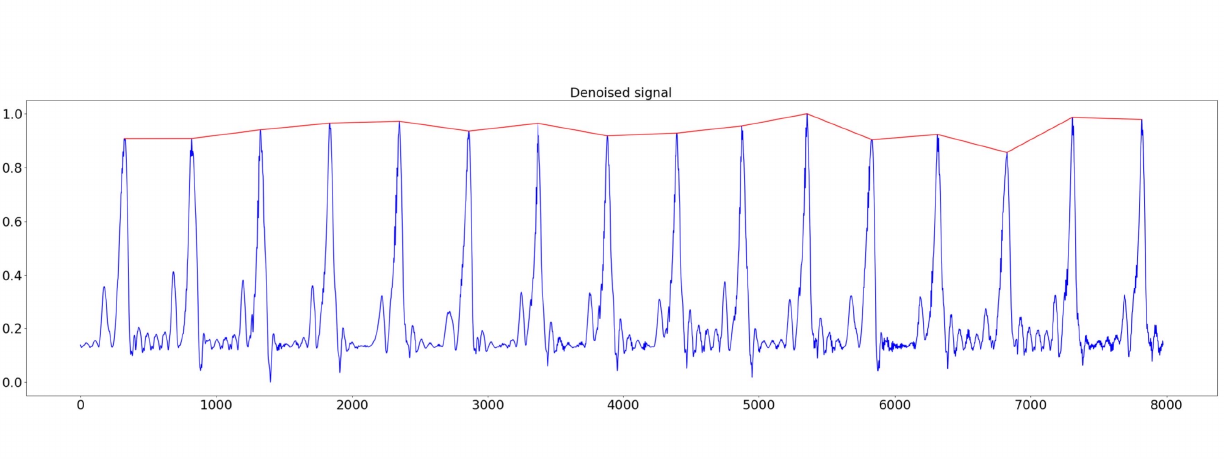}
\caption{Sub-figure 1. Denoised pre-cordial lead representative sample}
\label{tab:denoised-precordial}
\end{subfigure}

\caption{Comparison of a representative denoised limb lead signal and a representative pre-cordial lead signal.}
\label{fig:denoisecomaprisonfig}
\end{figure}

From a biological perspective, limb lead position is less defined amongst physicians \cite{Tung2021-ue-limb-lead-discrepencies}, and thus, since the MyoVista dataset (\ref{ssec:myovista}) was collected from multiple hospitals across multiple states, the positioning preferences between physicians could interfere with the model's ability to pinpoint predictive heuristics. Furthermore, the wavECG used to record the patient signals for the MyoVista dataset was designed to be able to identify left ventricular diastolic dysfunction or LVDD \cite{sengupta-etal-myovista}. LVDD is characterized by p-wave dispersion which is more identifiable in abnormalities in pre-cordial leads \cite{TAHA2016117}. Consequently, the wavECG hardware may have been implicitly designed to be more sensitive to the pre-cordial leads and thus the model may be able to identify more well-defined features in those leads as opposed to the limb leads. Computationally, following the application of the pre-processing procedure (\ref{ssec:pre-processing}) the denoised samples of our pre-cordial leads contain better-defined PQRST wave characteristics along with less noise overall when compared to the limb leads (Figure \ref{fig:denoisecomaprisonfig}). In particular, the P-wave is far more defined in each heartbeat of the pre-cordial lead (Figure \ref{tab:denoised-precordial}) which also provides further credence to the aforementioned argument that the wavECG is more sensitive to the pre-cordial leads.

\subsection{Error Analysis}
\label{ssec:erroranalysis}
\begin{figure}[htbp!]
\centering
\includegraphics[width=\columnwidth]{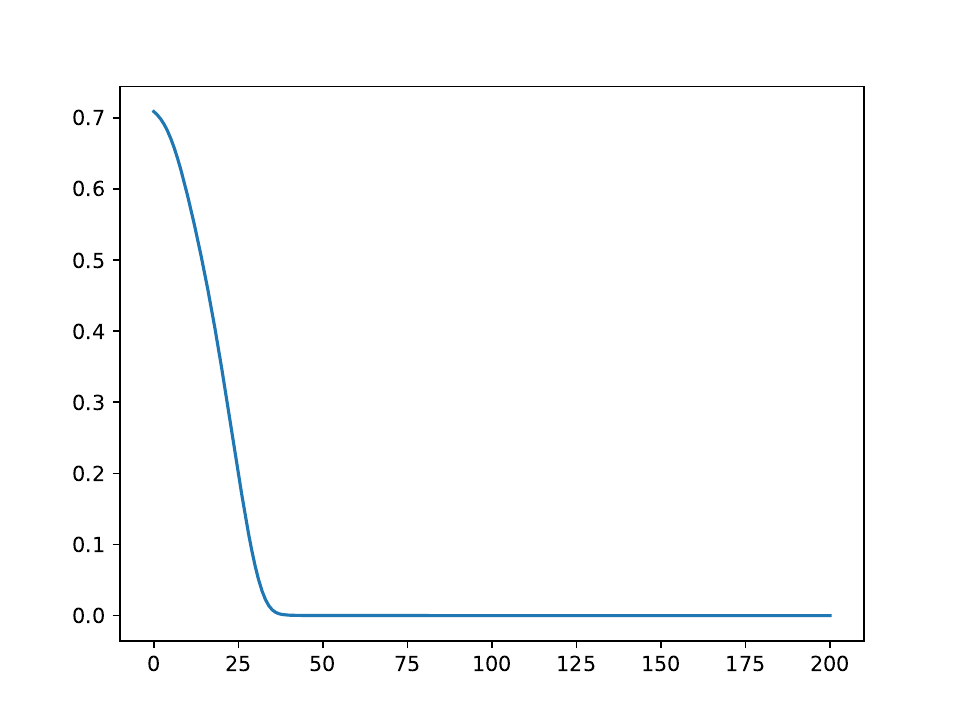}
\caption{A representative sample of a generated V5-lead signal. The V5 had the best mean predicted FD score during testing but fails to capture any PQRST complex features.}
\label{fig:v5}
\end{figure}

To understand the trends the model identifies in each lead and the inability of the model to reproduce all characteristics of the PQRST complex, we detailed manual error analysis of predictive examples from the lead best-predicted by the model (V5), the lead worst-predicted by the model (AVL), and a representative lead (V1). The lead for which the model performs the best for, V5, simply captures the downslope of the R peak and no other feature (Figure \ref{fig:v5}). 

\begin{figure}[htbp!]
\centering
\includegraphics[width=\columnwidth]{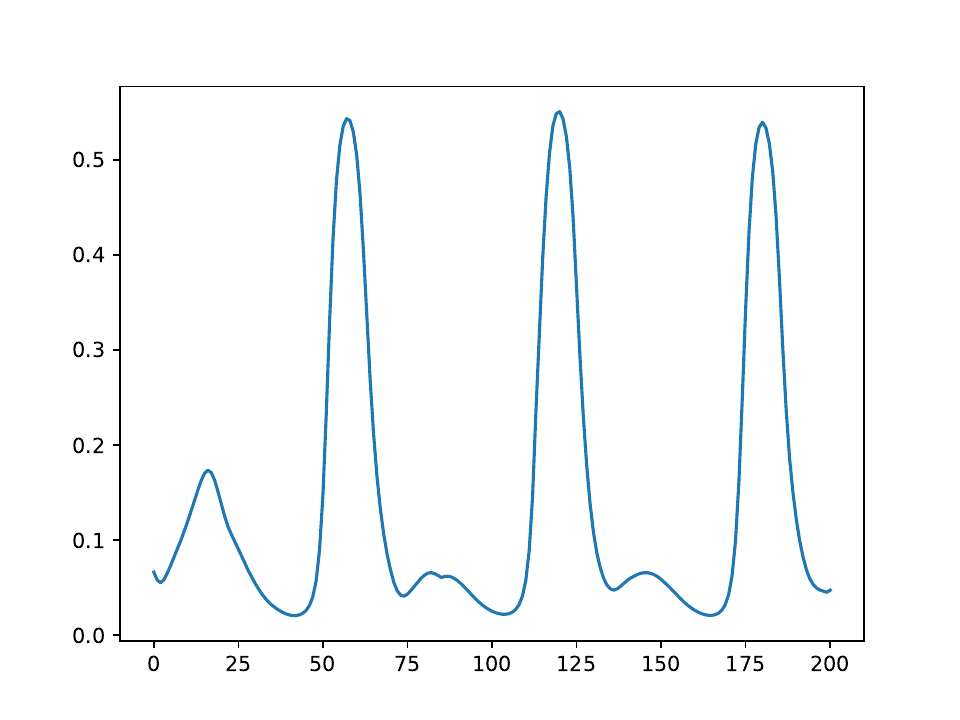}
\caption{A representative sample of a generated AVL-lead signal. The AVL-lead had the worst mean predicted FD score during testing and incorrectly generates heartbeats that have P-wave and T-wave peak magnitudes comparable to the R-peak.}
\label{fig:avl}
\end{figure}

This points to flaws in the FD score as the only metric to validate and test the model. Comparatively, the worst-predicted lead by the model, AVL, captures more features, notably a P-wave; however, it seems to predict multiple heartbeats instead of the intended single heartbeat (Figure \ref{fig:avl}). In this context, the depth dimension of the bidirectional LSTM generator (\ref{ssec:ganmodel}) appears to fail in differentiating between the heights of the P-wave peak, R-peak, and T-wave peak. 

\begin{figure}[htbp!]
\centering
\includegraphics[width=\columnwidth]{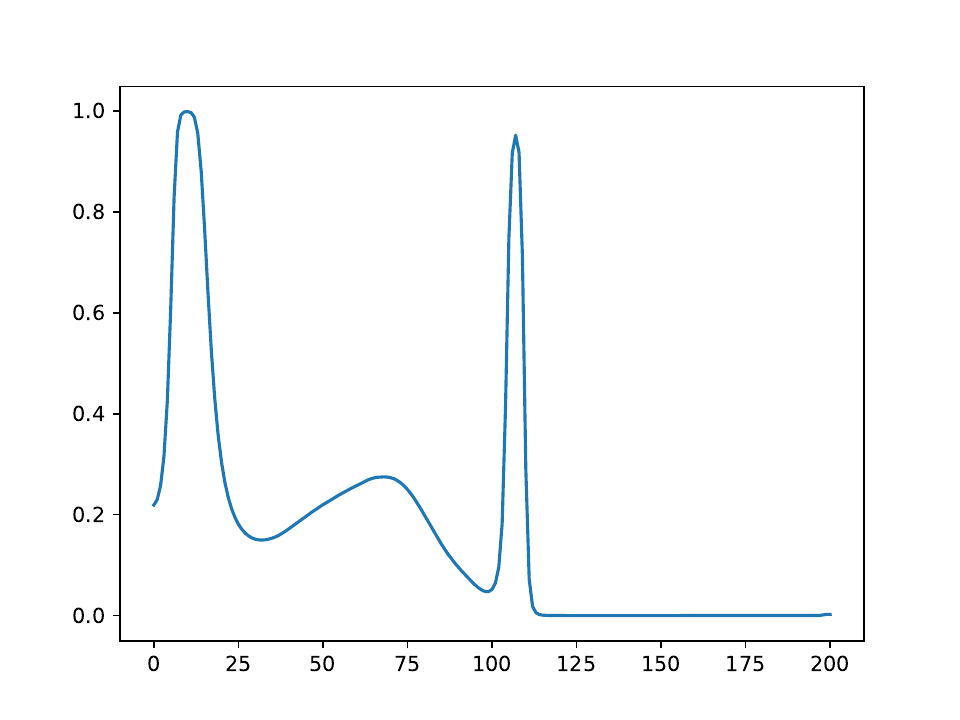}
\caption{A representative sample of a generated V1-lead signal. Out of all the generated signals, the V1 correctly preserves the most PQRST signals, most notably a dampened P-wave and a distinct R-peak.}
\label{fig:v1}
\end{figure}

The V1 lead, on the other hand, captures multiple features along with representing only one heartbeat despite having a middling FD score (Figure \ref{fig:v1}). The generated signal contains not only the down-slope of the previous R peak but a clear P-wave whose peak is accurately diminished compared to the predicted heartbeat's R peak; however, it still does not capture the T-wave. This can be attributed to the extensive zero-value padding that is sometimes appended to the end of each of the denoised signals (\ref{ssec:pre-processing}) in order to make all the heartbeats the same length for model input. Since the T-wave peaks' magnitudes are generally dampened in comparison to the P-wave peaks and R-peak \cite{COSTA2021957}, the model could mischaracterize the T-wave as signal noise and thus the generator perhaps never learned how to generate the T-wave. This could also point to flaws in our denoising procedure. Note I could not find a way to provide distributions of the categorized errors in a figure (see below) because the automatic identification of the components of the PQRST complexes themselves would require the training of an entirely new deep learning model; however, for publication, I will try to annotate as many of the figures as possible.

\subsection{Future Directions}
Because of the inability of the model to consistently and correctly characterize the necessary components of the PQRST complex for every lead, the model can be improved in future work in order to eventually identify conserved features that can be used as an input to a CVD-predictive model. This can be done by correcting the identified limitations mentioned in Section \ref{ssec:erroranalysis}. Foremost, the model is limited because we did not optimize the pre-processing algorithm—most notably the denoising algorithm—and thus a more advanced pre-processing algorithm could perhaps denoise the signal better such that the T-wave is not lost in signal generation. Although many previous generative and predictive models that take ECG signal inputs employ Daubechie discrete wavelet transforms (DWT) \cite{hazra-byun-2020-SynSigGAN, singh-pradhan-2021-denoising}, other models incorporate other methods. For example, \citet{zhu-etal-2019-bigridlstm} employs a variational autoencoder to correctly capture the distribution of data before performing a DWT-like procedure. Perhaps by training, validating, testing, and comparing multiple models that each take signals with a different pre-processing procedure performed, an optimal procedure can be identified and implemented so that more PQRST complex features are not misinterpreted as noise during model training. Furthermore, heartbeats with R-peak differentials that can be classified as statistical outliers can be eliminated in order to prevent extensive zero-padding to the segmented ECG signals which could also aid with T-wave preservation. This would also remove the need for the resampling we perform to reduce the input shape of the input layers of the bidirectional LSTM. By allocating time to tuning the hyperparameters of the bidirectional LSTM, perhaps the depth layers can effectively generate P-wave and T-wave peaks with the correct magnitudes. Finally, our analysis also hints at the flaws in Fréchet Distance as the evaluation metric for ECG signal generation and we will attempt to construct a novel evaluation metric. The hallmark of an accurately generated signal is the preservation of the PQRST complex and perhaps a metric that also measures the generated heartbeat's similarity to a standardized PQRST complex will prove successful. 
\label{ssec:futuredirections}
\section{Conclusion}
\label{sec:conclusion}

We present ECGNet that ultimately improves the accuracy of CVD-onset predictions by first improving the generation of a complete 12-lead ECG set from a single ECG lead signal input. Our generative adversarial network (GAN) model shows state-of-the-art accuracy in the generation of the pre-cordial lead signals—preserving distinct elements of the heartbeat's PQRST complex electrical signals. By applying cross-correlation analysis on ECGNet's GAN model's generated signals, latent features essential to the generation of accurate signals, and thus fundamental to the classification of CVD-onset, are identified. Note that the cross-correlation analysis and predictive CNN model were a part of Max's personal project and could not be completed during the semester's time frame; however, we chose to write about it hypothetically in the conclusion because it will be in the final publication and is essential in justifying the need for the project. Because of the identification of these hidden features, ECGNet allows cardiology to overcome the previous black box barriers that prevented the prediction of multiple-CVD targets. All our resources besides the dataset (due to HIPAA protections), but including models and source codes, are available through our open-source project at https://github.com/maxbagga/ScarletEagle1. Despite a decent performance, the GAN model still underperforms other state-of-the-art models in the generation of limb lead signals. We plan to tackle this challenge by collecting a second data set to train the limb lead generation with a different, non-MyoVista waveECG device. This will supplement the strengths of the waveECG in producing clearer pre-cordial lead signals while compensating for its tendencies to produce more noisy limb lead signals. Then, we will optimize pre-processing methods to give better limb lead inputs by experimenting with different denoising techniques and employing variational autoenconders to first capture the distribution of the data. Finally, we will aim to develop a new evaluation metric that considers the preservation of the PQRST complex in evaluating the accuracy of signal generation as the current metric employed—Fréchet Distance (FD)—only evaluates the location and orderings of signal points. Consequently, the leads with generated signals preserving the most PQRST complex elements often have middling FD scores which, with a novel method, will be rectified.

\section*{Acknowledgements}
We would like to thank Jinho Choi from Emory University, Naveena Yanamala from Carnegie Mellon Univeristy, and Partho Sengupta from Rutgers University for their help with the project.

\bibliography{custom}

\begin{thebibliography}{47}
\expandafter\ifx\csname natexlab\endcsname\relax\def\natexlab#1{#1}\fi

\bibitem[{Burguera(2019)}]{burguera-2019-QRS-detection}
Antoni Burguera. 2019.
\newblock \href {https://doi.org/10.1109/JBHI.2018.2792404} {Fast qrs detection
  and ecg compression based on signal structural analysis}.
\newblock volume~23, pages 123--131.

\bibitem[{Chiou et~al.(2021)Chiou, Syu, Wu, Lin, Yi, Lin, and
  Lin}]{chiou-etal-2021-ecg-predictions}
YA~Chiou, JY~Syu, SY~Wu, Lian-Yu Lin, Li~Tzu Yi, Ting-Tse Lin, and Shien-Fong
  Lin. 2021.
\newblock \href {https://doi.org/https://doi.org/10.1038/s41598-021-81374-6}
  {Electrocardiogram lead selection for intelligent screening of patients with
  systolic heart failure}.
\newblock In \emph{Scientific Reports Volume 11(1948)}, page~NA, London, United
  Kingdom. Nature Research.

\bibitem[{Chiu et~al.(2005)Chiu, Lin, and
  Liau}]{chiu-etal-2005-correlation-arrhythmia}
Chuang-Chien Chiu, Tong-Hong Lin, and Ben-Yi Liau. 2005.
\newblock \href {https://doi.org/10.4015/s1016237205000238} {Using correlation
  coefficient in ecg waveform for arrhythmia detection}.
\newblock \emph{Biomedical Engineering: Applications, Basis and
  Communications}, 17(03):147–152.

\bibitem[{Costa et~al.(2021)Costa, Winkert, Manhães, and
  Teixeira}]{COSTA2021957}
Renan Costa, Thaís Winkert, Aline Manhães, and João~Paulo Teixeira. 2021.
\newblock \href {https://doi.org/https://doi.org/10.1016/j.procs.2021.01.252}
  {Qrs peaks, p and t waves identification in ecg}.
\newblock \emph{Procedia Computer Science}, 181:957--964.
\newblock CENTERIS 2020 - International Conference on ENTERprise Information
  Systems / ProjMAN 2020 - International Conference on Project MANagement /
  HCist 2020 - International Conference on Health and Social Care Information
  Systems and Technologies 2020, CENTERIS/ProjMAN/HCist 2020.

\bibitem[{Craven et~al.(2017)Craven, McGinley, Kilmartin, Glavin, and
  Jones}]{craven-etal-2017-ECG-proof}
Darren Craven, Brian McGinley, Liam Kilmartin, Martin Glavin, and Edward Jones.
  2017.
\newblock \href {https://doi.org/10.1109/JBHI.2016.2531182} {Adaptive
  dictionary reconstruction for compressed sensing of ecg signals}.
\newblock volume~21, pages 645--654.

\bibitem[{Dawber et~al.(1952)Dawber, Kannel, Love, and
  Streeper}]{dawber-etal-1952-ecg-traditional-use}
TR~Dawber, WB~Kannel, DE~Love, and RB~Streeper. 1952.
\newblock \href {https://doi.org/https://doi.org/10.1161/01.CIR.5.4.559} {The
  electrocardiogram in heart disease detection a comparison of the multiple and
  single lead procedures}.
\newblock In \emph{Circulation Volume V}, pages 559--566, NA. Lippincott
  Williams \& Wilkins.

\bibitem[{Drew et~al.(2002)Drew, Pelter, Brodnick, Yadav, Dempel, and
  Adams}]{drew-etal-2002-interpolation}
Barbara~J. Drew, Michele~M. Pelter, Donald~E. Brodnick, Anil~V. Yadav, Debbie
  Dempel, and Mary~G. Adams. 2002.
\newblock \href {https://doi.org/https://doi.org/10.1054/jelc.2002.37150}
  {Comparison of a new reduced lead set ecg with the standard ecg for
  diagnosing cardiac arrhythmias and myocardial ischemia}.
\newblock In \emph{Journal of Electrocardiology Volume 35(4B)}, pages 13--21,
  NA. Elsevier.

\bibitem[{Dutta et~al.(2020)Dutta, Batabyal, Basu, and
  Acton}]{dutta-etal-2020-CNN3}
Aniruddha Dutta, Tamal Batabyal, Meheli Basu, and Scott~T. Acton. 2020.
\newblock \href {https://doi.org/https://doi.org/10.1016/j.eswa.2020.113408}
  {An efficient convolutional neural network for coronary heart disease
  prediction}.
\newblock \emph{Expert Systems with Applications}, 159:113408.

\bibitem[{Ebrahimi et~al.(2020)Ebrahimi, Loni, Daneshtalab, and
  Gharehbaghi}]{ebrahimi-etal-2020-cardiologist-shortcomings}
Zahra Ebrahimi, Mohammad Loni, Masoud Daneshtalab, and Arash Gharehbaghi. 2020.
\newblock \href {https://doi.org/https://doi.org/10.1016/j.eswax.2020.100033}
  {A review on deep learning methods for ecg arrhythmia classification}.
\newblock In \emph{Expert Systems with Applications: X Volume 7}, page~NA, NA.
  Elsevier.

\bibitem[{Grande-Fidalgo et~al.(2021)Grande-Fidalgo, Calpe, Redón,
  Millán-Navarro, and Soria-Olivas}]{grande-fildago-etal-2021-ECG-proof}
Alejandro Grande-Fidalgo, Javier Calpe, Mónica Redón, Carlos Millán-Navarro,
  and Emilio Soria-Olivas. 2021.
\newblock \href {https://doi.org/10.3390/s21165542} {Lead reconstruction using
  artificial neural networks for ambulatory ecg acquisition}.
\newblock volume~21.

\bibitem[{Hazra and Byun(2020)}]{hazra-byun-2020-SynSigGAN}
Debapriaya Hazra and Yung-Cheol Byun. 2020.
\newblock \href {https://doi.org/https://doi.org/10.3390/biology9120441}
  {Synsiggan: Generative adversarial networks for synthetic biomedical signal
  generation}.
\newblock In \emph{Biology Volume 9(12)}, page~NA, NA. Multidisciplinary
  Digital Publishing Institute.

\bibitem[{He et~al.(2018)He, Sun, Rong, Wang, and
  Zhang}]{he-etal-2018-heartbeat-classification}
Jinyuan He, Le~Sun, Jia Rong, Hua Wang, and Yanchun Zhang. 2018.
\newblock \href {https://doi.org/10.1371/journal.pone.0206593} {A pyramid-like
  model for heartbeat classification from ecg recordings}.
\newblock volume~13.

\bibitem[{Hochreiter and Schmidhuber(1997)}]{hochreiter-schmidhuber-1997-lstm}
Sepp Hochreiter and Jürgen Schmidhuber. 1997.
\newblock \href {https://doi.org/10.1162/neco.1997.9.8.1735} {{Long Short-Term
  Memory}}.
\newblock \emph{Neural Computation}, 9(8):1735--1780.

\bibitem[{Hsu and Wu(2014)}]{hsu-wu-2014-three-lead-inter}
Chih-Hao Hsu and Sau-Hsuan Wu. 2014.
\newblock \href {https://doi.org/10.1109/ICC.2014.6883866} {Robust signal
  synthesis of the 12-lead ecg using 3-lead wireless ecg systems}.
\newblock In \emph{2014 IEEE International Conference on Communications (ICC)},
  pages 3517--3522.

\bibitem[{Iwana et~al.(2019)Iwana, Kuroki, and
  Uchida}]{iwana-etal-2019-softmax}
Brian~Kenji Iwana, Ryohei Kuroki, and Seiichi Uchida. 2019.
\newblock \href {https://doi.org/10.1109/ICCVW.2019.00513} {Explaining
  convolutional neural networks using softmax gradient layer-wise relevance
  propagation}.
\newblock In \emph{2019 IEEE/CVF International Conference on Computer Vision
  Workshop (ICCVW)}, pages 4176--4185.

\bibitem[{Kachuee et~al.(2018)Kachuee, Fazeli, and
  Sarrafzadeh}]{kachuee-etal-2018-heartbeat-classification-mathematical}
Mohammad Kachuee, Shayan Fazeli, and Majid Sarrafzadeh. 2018.
\newblock \href {https://doi.org/10.1109/ICHI.2018.00092} {Ecg heartbeat
  classification: A deep transferable representation}.
\newblock In \emph{2018 IEEE International Conference on Healthcare Informatics
  (ICHI)}, pages 443--444.

\bibitem[{Kiranyaz et~al.(2021)Kiranyaz, Avci, Abdeljaber, Ince, Gabbouj, and
  Inman}]{serkan-etal-2021-1dCNN}
Serkan Kiranyaz, Onur Avci, Osama Abdeljaber, Turker Ince, Moncef Gabbouj, and
  Daniel~J. Inman. 2021.
\newblock \href {https://doi.org/https://doi.org/10.1016/j.ymssp.2020.107398}
  {1d convolutional neural networks and applications: A survey}.
\newblock \emph{Mechanical Systems and Signal Processing}, 151:107398.

\bibitem[{Kiranyaz et~al.(2019)Kiranyaz, Ince, Abdeljaber, Avci, and
  Gabbouj}]{kiranyaz-etal-2019-1dCnn}
Serkan Kiranyaz, Turker Ince, Osama Abdeljaber, Onur Avci, and Moncef Gabbouj.
  2019.
\newblock \href {https://doi.org/10.1109/ICASSP.2019.8682194} {1-d
  convolutional neural networks for signal processing applications}.
\newblock In \emph{ICASSP 2019 - 2019 IEEE International Conference on
  Acoustics, Speech and Signal Processing (ICASSP)}, pages 8360--8364.

\bibitem[{Lee et~al.(2022)Lee, Gommers, Wohlfahrt, Wasilewski, O'Leary,
  Nahrstaedt, Sauvé, Agrawal, Pelt, Oliveira, Arildsen, Clauss, Yu, Brett,
  Pelletier, SylvainLan, Tricoli, Choudhary, asnt, Smith, 0-tree, Reczey,
  Goldberg, Goertzen, Laszuk, ElConno, Antonello, Mandula, jakirkham, and
  Dan}]{gregory-etal-2022-pywt}
Gregory Lee, Ralf Gommers, Kai Wohlfahrt, Filip Wasilewski, Aaron O'Leary,
  Holger Nahrstaedt, Alexandre Sauvé, Ankit Agrawal, Daniel~M. Pelt, Helder
  Oliveira, Thomas Arildsen, Christian Clauss, Frank Yu, Matthew Brett, Michel
  Pelletier, SylvainLan, Daniele Tricoli, Saket Choudhary, asnt, Arfon Smith,
  0-tree, Balint Reczey, Corey Goldberg, Daniel Goertzen, Dawid Laszuk,
  ElConno, Jacopo Antonello, Jakub Mandula, jakirkham, and Jonathan Dan. 2022.
\newblock \href {https://doi.org/10.5281/zenodo.7086189} {Pywavelets/pywt:
  v1.4.1}.

\bibitem[{Lee et~al.(2017{\natexlab{a}})Lee, Lee, Kwon, Kim, and
  Park}]{lee-etal-2017-single-patch}
Hong~J. Lee, Dong~S. Lee, Hyun~B. Kwon, Do~Y. Kim, and Kwang~S. Park.
  2017{\natexlab{a}}.
\newblock \href {https://doi.org/10.3414/me16-01-0067} {Reconstruction of
  12-lead ecg using a single-patch device}.
\newblock volume~56, page 319–327.

\bibitem[{Lee et~al.(2017{\natexlab{b}})Lee, Kim, and
  Shin}]{lee-etal-2017-drowsiness}
Jaewon Lee, Jinwoo Kim, and Miyoung Shin. 2017{\natexlab{b}}.
\newblock \href {https://doi.org/https://doi.org/10.1016/j.procs.2017.10.083}
  {Correlation analysis between electrocardiography (ecg) and
  photoplethysmogram (ppg) data for driver’s drowsiness detection using noise
  replacement method}.
\newblock \emph{Procedia Computer Science}, 116:421--426.
\newblock Discovery and innovation of computer science technology in artificial
  intelligence era: The 2nd International Conference on Computer Science and
  Computational Intelligence (ICCSCI 2017).

\bibitem[{Li and Boulanger(2020)}]{li-boulanger-2020-ecg-dl}
Hongzu Li and Pierre Boulanger. 2020.
\newblock \href {https://doi.org/https://doi.org/10.3390/s20051461} {A survey
  of heart anomaly detection using ambulatory electrocardiogram (ecg)}.
\newblock In \emph{Sensors Volume 20(5)}, page~NA, NA. Multidisciplinary
  Digital Publishing Institute.

\bibitem[{Mir and Singh(2021)}]{haroon-singh-2021-denoise}
Haroon~Yousuf Mir and Omkar Singh. 2021.
\newblock \href {https://doi.org/10.1080/03091902.2021.1955032} {Ecg denoising
  and feature extraction techniques – a review}.
\newblock volume~45, pages 672--684. Taylor \& Francis.
\newblock PMID: 34463593.

\bibitem[{Oosterom(2002)}]{oosterom-2002-solid-angle}
A.~van Oosterom. 2002.
\newblock \href {https://doi.org/https://doi.org/10.1054/jelc.2002.37176}
  {Solidifying the solid angle}.
\newblock In \emph{Journal of Electrocardiology Volume 35(4)}, pages 181--192,
  NA. Elsevier.

\bibitem[{Podobnik and Stanley(2008)}]{podobnik-stanley-2008-cross-correlation}
Boris Podobnik and H.~Eugene Stanley. 2008.
\newblock \href {https://doi.org/10.1103/PhysRevLett.100.084102} {Detrended
  cross-correlation analysis: A new method for analyzing two nonstationary time
  series}.
\newblock \emph{Phys. Rev. Lett.}, 100:084102.

\bibitem[{Ramli and Ahmad(2003)}]{ramli-ahmad-2003-cross-correlation}
A.B. Ramli and P.A. Ahmad. 2003.
\newblock \href {https://doi.org/10.1109/NCTT.2003.1188342} {Correlation
  analysis for abnormal ecg signal features extraction}.
\newblock In \emph{4th National Conference of Telecommunication Technology,
  2003. NCTT 2003 Proceedings.}, pages 232--237.

\bibitem[{Roth et~al.(2020)Roth, Mensah, Johnson, Addolorato, Ammirati,
  Baddour, Barengo, Beaton, Benjamin, Benzinger, Bonny, Brauer, Brodmann,
  Cahill, Carapetis, Catapano, Chugh, Cooper, Coresh, Criqui, DeCleene, Eagle,
  Emmons-Bell, Feigin, Fernández-Solà, Fowkes, Gakidou, Grundy, He, Howard,
  Hu, Inker, Karthikeyan, Kassebaum, Koroshetz, Lavie, Lloyd-Jones, Lu,
  Mirijello, Temesgem, Mokdad, Moran, Muntner, Narula, Neal, Ntsekhe, Moraes~de
  Oliveira, Otto, Owolabi, Pratt, Rajagopalan, Reitsma, Ribeiro, Rigotti,
  Rodgers, Sable, Shakil, Sliwa-Hahnle, Stark, Sundström, Timpel, Tleyjeh,
  Valgimigli, Vos, Whelton, Yacoub, Zuhlke, Murray, and
  Fuster}]{roth-etal-2020-CVD-mortality}
GA~Roth, GA~Mensah, CO~Johnson, G~Addolorato, E~Ammirati, LM~Baddour,
  NC~Barengo, AZ~Beaton, EJ~Benjamin, CP~Benzinger, A~Bonny, M~Brauer,
  M~Brodmann, TJ~Cahill, J~Carapetis, AL~Catapano, SS~Chugh, LT~Cooper,
  J~Coresh, M~Criqui, N~DeCleene, KA~Eagle, S~Emmons-Bell, VL~Feigin,
  J~Fernández-Solà, G~Fowkes, E~Gakidou, SM~Grundy, FJ~He, G~Howard, F~Hu,
  L~Inker, G~Karthikeyan, N~Kassebaum, W~Koroshetz, C~Lavie, D~Lloyd-Jones,
  HS~Lu, A~Mirijello, AM~Temesgem, A~Mokdad, AE~Moran, P~Muntner, J~Narula,
  B~Neal, M~Ntsekhe, G~Moraes~de Oliveira, C~Otto, M~Owolabi, M~Pratt,
  S~Rajagopalan, M~Reitsma, ALP Ribeiro, N~Rigotti, A~Rodgers, C~Sable,
  S~Shakil, K~Sliwa-Hahnle, B~Stark, J~Sundström, P~Timpel, IM~Tleyjeh,
  M~Valgimigli, T~Vos, PK~Whelton, M~Yacoub, L~Zuhlke, C~Murray, and V~Fuster.
  2020.
\newblock \href {https://doi.org/https://doi.org/10.1016/j.jacc.2020.11.010}
  {Global burden of cardiovascular diseases and risk factors, 1990–2019}.
\newblock In \emph{Journal of the American College of Cardiology Volume 76
  (25)}, pages 2982--3021, NA. Elsevier.

\bibitem[{Sajja and Kalluri(2020)}]{sajja-kalluri-2020-CNN-superiority}
Tulasi~Krishna Sajja and Hemantha~Kumar Kalluri. 2020.
\newblock \href {https://doi.org/10.18280/ria.340510} {A deep learning method
  for prediction of cardiovascular disease using convolutional neural network}.
\newblock \emph{Revue d'Intelligence Artificielle}, 34(5):601–606.

\bibitem[{Schuster and Paliwal(1997)}]{schuster-paliwal-1997-bigrid}
M.~Schuster and K.K. Paliwal. 1997.
\newblock \href {https://doi.org/10.1109/78.650093} {Bidirectional recurrent
  neural networks}.
\newblock \emph{IEEE Transactions on Signal Processing}, 45(11):2673--2681.

\bibitem[{Sengupta et~al.(2018{\natexlab{a}})Sengupta, Kulkarni, and
  Narula}]{sengupta-etal-myovista}
Partho~P. Sengupta, Hemant Kulkarni, and Jagat Narula. 2018{\natexlab{a}}.
\newblock \href {https://doi.org/10.1016/j.jacc.2018.02.024} {Prediction of
  abnormal myocardial relaxation from signal processed surface ecg}.
\newblock \emph{Journal of the American College of Cardiology},
  71(15):1650--1660.

\bibitem[{Sengupta et~al.(2018{\natexlab{b}})Sengupta, Kulkarni, and
  Narula}]{Sengupta-etal-2018-wavecg}
Partho~P. Sengupta, Hemant Kulkarni, and Jagat Narula. 2018{\natexlab{b}}.
\newblock \href {https://doi.org/10.1016/j.jacc.2018.02.024} {Prediction of
  abnormal myocardial relaxation from signal processed surface ecg}.
\newblock \emph{Journal of the American College of Cardiology},
  71(15):1650--1660.

\bibitem[{Seo et~al.(2022)Seo, Yoon, Joo, and
  Nam}]{hyo-chang-etal-2022-similar-one}
Hyo-Chang Seo, Gi-Won Yoon, Segyeong Joo, and Gi-Byoung Nam. 2022.
\newblock \href {https://doi.org/https://doi.org/10.1016/j.cmpb.2022.106858}
  {Multiple electrocardiogram generator with single-lead electrocardiogram}.
\newblock volume 221, page 106858.

\bibitem[{Shankar et~al.(2020)Shankar, Kumar, Devagade, Karanth, and
  Rohitaksha}]{shankar-etal-2020-CNN2}
VirenViraj Shankar, Varun Kumar, Umesh Devagade, Vinay Karanth, and
  K.~Rohitaksha. 2020.
\newblock \href {https://doi.org/10.1007/s42979-020-0097-6} {Heart disease
  prediction using cnn algorithm}.
\newblock \emph{SN Computer Science}, 1(3).

\bibitem[{Sherstinsky(2020)}]{Sherstinsky-2020-RNN}
Alex Sherstinsky. 2020.
\newblock \href {https://doi.org/https://doi.org/10.1016/j.physd.2019.132306}
  {Fundamentals of recurrent neural network (rnn) and long short-term memory
  (lstm) network}.
\newblock \emph{Physica D: Nonlinear Phenomena}, 404:132306.

\bibitem[{Singh and Pradhan(2021)}]{singh-pradhan-2021-denoising}
Pratik Singh and Gayadhar Pradhan. 2021.
\newblock \href {https://doi.org/10.1109/TCBB.2020.2976981} {A new ecg
  denoising framework using generative adversarial network}.
\newblock volume~18, pages 759--764.

\bibitem[{Sohn et~al.(2020)Sohn, Yang, Lee, and
  Kim}]{sohn-etal-2020-three-lead-dl}
Jangjay Sohn, Seungman Yang, Yunseo Lee, Joonnyong~Ku, and Hee~Chan Kim. 2020.
\newblock \href {https://doi.org/https://doi.org/10.3390/s20113278}
  {Reconstruction of 12-lead electrocardiogram from a three-lead patch-type
  device using a lstm network}.
\newblock In \emph{Sensors Volume 20(11)}, page~NA, NA. Multidisciplinary
  Digital Publishing Institute.

\bibitem[{Stracina et~al.(2022)Stracina, Ronzhina, Redina, and
  Novakova}]{stracina-etal-2021-clinical}
Tibor Stracina, Marina Ronzhina, Richard Redina, and Marie Novakova. 2022.
\newblock \href {https://doi.org/https://doi.org/10.3389/fphys.2022.867033}
  {Golden standard or obsolete method? review of ecg applications in clinical
  and experimental context}.
\newblock In \emph{Frontiers in Physiology Volume}, page~NA, NA. Frontiers
  Media S.A.

\bibitem[{Taha et~al.(2016)Taha, Sayed, Saad, and Samir}]{TAHA2016117}
Tamer Taha, Khaled Sayed, Mohamad Saad, and Mohammed Samir. 2016.
\newblock \href {https://doi.org/https://doi.org/10.1016/j.ehj.2015.01.002}
  {How accurate can electrocardiogram predict left ventricular diastolic
  dysfunction?}
\newblock \emph{The Egyptian Heart Journal}, 68(2):117--123.

\bibitem[{Toyoshima et~al.(1958)Toyoshima, Kato, Isobe, Kutsuna, Nagaya, and
  Saruhashi}]{toyoshima-etal-1958-original-reconstruction}
Hideo Toyoshima, Hiroshi Kato, Takehiko Isobe, Yoshio Kutsuna, Teruo Nagaya,
  and Yoshiko Saruhashi. 1958.
\newblock \href {https://doi.org/https://doi.org/10.1016/0002-8703(58)90229-1}
  {Electrocardiogram and vectorcardiogram reconstruction and its application to
  clinical diagnosis of myocardial infarction}.
\newblock In \emph{American Heart Journal Volume 56(2)}, pages 165--194, NA.
  Elsevier.

\bibitem[{Tung(2021)}]{Tung2021-ue-limb-lead-discrepencies}
Robert~T Tung. 2021.
\newblock Electrocardiographic limb leads placement and its clinical
  implication: Two cases of electrocardiographic illustrations.
\newblock \emph{Kans J Med}, 14:229--230.

\bibitem[{Ulloa-Cerna et~al.(2022)Ulloa-Cerna, Jing, Pfeifer, Raghunath, Ruhl,
  Rocha, Leader, Zimmerman, Lee, Steinhubl, Haggerty, and
  Fornwalt}]{ulloa-cerna-2022-rECHOmmend}
Alvaro Ulloa-Cerna, Linyuan Jing, John Pfeifer, Sushravya Raghunath, Jeffrey
  Ruhl, Daniel Rocha, Joseph Leader, Noah Zimmerman, Greg Lee, Christopher
  Steinhubl, Steven~Good, Christopher Haggerty, and Brandon Fornwalt. 2022.
\newblock \href
  {https://doi.org/https://doi.org/10.1161/CIRCULATIONAHA.121.057869}
  {rechommend: An ecg-based machine learning approach for identifying patients
  at increased risk of undiagnosed structural heart disease detectable by
  echocardiography}.
\newblock In \emph{Circulation Volume 146(1)}, pages 36--47, NA. Lippincott
  Williams \& Wilkins.

\bibitem[{Van~Rossum and Drake~Jr(1995)}]{van-drake-1995-python}
Guido Van~Rossum and Fred~L Drake~Jr. 1995.
\newblock \emph{Python reference manual}.
\newblock Centrum voor Wiskunde en Informatica Amsterdam.

\bibitem[{Weimann and Conrad(2021)}]{weimann-conrad-2021-CNN-transfer-learning}
Kuba Weimann and Tim~O. Conrad. 2021.
\newblock \href {https://doi.org/10.1038/s41598-021-84374-8} {Transfer learning
  for ecg classification}.
\newblock \emph{Scientific Reports}, 11(1).

\bibitem[{Wu et~al.(2021)Wu, Lu, Yang, and Wong}]{wu-etal-2021-CNN1}
Mengze Wu, Yongdi Lu, Wenli Yang, and Shen~Yuong Wong. 2021.
\newblock \href {https://doi.org/10.3389/fncom.2020.564015} {A study on
  arrhythmia via ecg signal classification using the convolutional neural
  network}.
\newblock \emph{Frontiers in Computational Neuroscience}, 14.

\bibitem[{Zhou et~al.(2021)Zhou, Guo, Wang, Wu, Li, Yao, Fang, Yang, Cao, and
  Cui}]{zhou-etal-2021-functional-CVD}
Liye Zhou, Zhifei Guo, Bijue Wang, Yongqin Wu, Zhi Li, Hongmei Yao, Ruiling
  Fang, Haitao Yang, Hongyan Cao, and Yuehua Cui. 2021.
\newblock \href {https://doi.org/https://doi.org/10.3389/fgene.2021.652315}
  {Risk prediction in patients with heart failure with preserved ejection
  fraction using gene expression data and machine learning.}
\newblock In \emph{Frontiers in Genetics}, page~NA, NA. Frontiers Media S.A.

\bibitem[{Zhu et~al.(2019)Zhu, Ye, Fu, Liu, and
  Shen}]{zhu-etal-2019-bigridlstm}
Fei Zhu, Fei Ye, Yuchen Fu, Quan Liu, and Bairong Shen. 2019.
\newblock \href {https://doi.org/https://doi.org/10.1038/s41598-019-42516-z}
  {Electrocardiogram generation with a bidirectional lstm-cnn generative
  adversarial network}.
\newblock In \emph{Scientific Reports Volume 9 (6374)}, page N/A, London,
  United Kingdom. Nature Research.

\bibitem[{Zhu et~al.(2017)Zhu, Park, Isola, and
  Efros}]{Zhu-etal-2017-loss-function}
Jun-Yan Zhu, Taesung Park, Phillip Isola, and Alexei~A. Efros. 2017.
\newblock \href {https://doi.org/10.1109/ICCV.2017.244} {Unpaired
  image-to-image translation using cycle-consistent adversarial networks}.
\newblock In \emph{2017 IEEE International Conference on Computer Vision
  (ICCV)}, pages 2242--2251.

\end{thebibliography}

\cleardoublepage\appendix

\end{document}